\pdfoutput=1
\documentclass[journal]{IEEEtran}

\usepackage{booktabs}
\usepackage{float}
\usepackage[colorlinks,linkcolor=blue]{hyperref}
\usepackage[latin1]{inputenc}
\usepackage{amsmath}
\usepackage{graphicx}
\usepackage{amssymb}
\usepackage{amsthm}
\usepackage{verbatim}
\usepackage{epsfig}
\usepackage[labelfont=bf]{caption}
\usepackage{subcaption}
\usepackage{enumerate}
\usepackage{epstopdf}

\usepackage{stfloats}
\usepackage{amssymb}
\usepackage{amsthm}
\usepackage{graphicx}
\usepackage{float}
\usepackage{amsmath}
\usepackage{amsmath,bm}
\usepackage{color}
\usepackage{multirow}
\usepackage{mathrsfs}
\usepackage{lineno,hyperref}
\usepackage{lipsum}
\usepackage{stfloats}
\usepackage{titletoc}
\usepackage{dsfont}
\usepackage{mathtools}

\newtheorem{definition}{Definition}

\newtheorem{lemma}{Lemma}
\newtheorem{remark}{Remark}

\def\begquo{\begin{quote}}
	\def\endquo{\end{quote}}
\def\begequarr{\begin{eqnarray}}
\def\endequarr{\end{eqnarray}}
\def\begequarrs{\begin{eqnarray*}}
	\def\endequarrs{\end{eqnarray*}}
\def\begarr{\begin{array}}
	\def\endarr{\end{array}}
\def\begequ{\begin{equation}}
\def\endequ{\end{equation}}
\def\lab{\label}
\def\begdes{\begin{description}}
	\def\enddes{\end{description}}
\def\begenu{\begin{enumerate}}
	\def\begite{\begin{itemize}}
		\def\endite{\end{itemize}}
	\def\endenu{\end{enumerate}}

\def\lef[{\left[\begin{array}}
	\def\rig]{\end{array}\right]}

\def\begcen{\begin{center}}
	\def\endcen{\end{center}}
\def\begrem{\begin{remark}\rm}
	\def\endrem{\end{remark}}
\def\begdef{\begin{definition}}
	\def\enddef{\end{definition}}
\def\begpro{\begin{propositionosition}}
	\def\endpro{\end{propositionosition}}
\def\begfac{\begin{fact}}
	\def\endfac{\end{fact}}
\def\begass{\begin{assumptionption}}
	\def\endass{\end{assumptionption}}

\def\begmat#1{\begin{bmatrix}#1\end{bmatrix}}
\def\begali#1{\begin{align}{#1}\end{align}}
\def\begalis#1{\begin{align*}{#1}\end{align*}}

\def\liminf{\lim_{t \to \infty}}

\def\L2e{{\cal L}_{2e}}

\def\rea{\mathbb{R}}

\def\adj{\mbox{adj}}

\usepackage{color}

\newcommand{\mrm}{\mathrm}

\usepackage[prependcaption,colorinlistoftodos]{todonotes}

\title{Dynamic state and parameter estimation in multi-machine power systems - Experimental demonstration using real-world PMU-measurements}
\author{Nicolai Lorenz-Meyer, René Suchantke, Johannes Schiffer~\IEEEmembership{Member,~IEEE}
\thanks{N. Lorenz-Meyer is with Brandenburg University of Technology Cottbus-Senftenberg, 03046 Cottbus, Germany (e-mail: lorenz-meyer@b-tu.de).}
\thanks{R. Suchantke is with 50Hertz Transmission GmbH, 10557 Berlin, Germany (e-mail: rene.suchantke@50hertz.com).}
\thanks{J. Schiffer is with Brandenburg University of Technology Cottbus-Senftenberg, 03046 Cottbus, Germany and Fraunhofer Research Institution for Energy Infrastructures and Geothermal Systems (IEG), 03046 Cottbus, Germany (e-mail: schiffer@b-tu.de).}
}

\begin{document}
\maketitle
\begin{abstract}
Dynamic state and parameter estimation (DSE) plays a key role for reliably monitoring and operating future, power-electronics-dominated power systems. While DSE is a very active research field, experimental applications of proposed algorithms to real-world systems remain scarce. This motivates the present paper, in which we demonstrate the effectiveness of a DSE algorithm 
 previously presented by parts of the authors with real-world data collected by a Phasor Measurement Unit (PMU) at a substation close to a power plant within the extra-high voltage grid of Germany. 
To this end, at first we derive a suitable mapping of the real-world PMU-measurements recorded at a substation close to the power plant to the terminal bus of the power plants' synchronous generator (SG). This mapping considers the high-voltage (HV) transmission line, the tap-changing transformer and the auxiliary system of the power plant.
Next, we introduce several practically motivated extensions to the estimation algorithm, which significantly improve its practical performance with real-world measurements. 
Finally, we successfully validate the algorithm experimentally in an auto- as well as a cross-validation. 
\end{abstract}

\begin{IEEEkeywords}
	Real-world PMU-measurements, dynamic state and parameter estimation, experimental demonstration, power grid monitoring, power system operation, phasor measurements, synchronous generator.
\end{IEEEkeywords}

\section{Introduction}

\subsection{Motivation and existing literature} 
\IEEEPARstart{D}{ynamic} monitoring of the transient behavior of power systems is increasingly important due to the major changes implemented as a consequence of the world-wide energy transition. Vastly increasing amounts of renewable, power-electronics interfaced energy sources are deployed in different hierarchical levels of the power systems. Meanwhile, more complex loads and novel demand-response technologies are introduced \cite{winter_pushing_2015}. This yields higher and reverse power flows, causing faster and more volatile dynamics of the overall system and leading to an operation closer to the stability limit \cite{milano_foundations_2018}.
Thus, dependable methods for DSE are of utmost importance for the reliable operation of future power systems \cite{zhao_et_al_power_2021}.
Moreover, accurate knowledge of the dynamic parameters of power system components and initial conditions are essential for accurate simulation within the framework of dynamic security assessment \cite{vittal_et_al_next_2011}. This is an essential topic for 50Hertz Transmission GmbH in order to setup reliable dynamic simulations. 

Enabled by the expanding deployment of Phasor Measurement Units (PMUs) and advanced communication infrastructure, novel DSE designs are facilitated and the area of DSE has (again) become an active field of research \cite{zhao_et_al_power_2021}. Most works presented in the literature focus on Kalman filter (KF-)based methods, which include extended KFs (EKFs) \cite{paul_dynamic_2018}, unscented KFs (UKFs) \cite{wang_alternative_2012, valverde_unscented_2010} and particle filters \cite{emami_particle_2015, cui_particle_2015}. In recent years, efforts have been made to propose methods based on state observation techniques for nonlinear systems and providing rigorous convergence guarantees. A robust observer with known inputs is proposed in \cite{nugroho_robust_2020} and in \cite{taha_risk_2018} a sliding mode observer for a risk mitigation strategy is developed. In \cite{qi_comparing_2018} a cubature KF and a nonlinear observer are proposed and compared. 

Yet, in all above mentioned works the developed methods are only validated using simulation studies. In fact, experimental demonstration of DSE methods employing PMU-measurements from real-world power systems are only very rarely reported in the literature. In \cite{simendic_-field_2005} a real-time distribution state estimator is presented for steady state conditions. The proposed method is successfully verified using Supervisory Control and Data Acquisition (SCADA) data from the Distribution Utility Elektrovojvodina in Serbia. The same method is tested in real-life operation of the Distribution Utility Guizhou Power Corporation in China in \cite{katic_field_2013}.


\subsection{Contributions}
Motivated by the aforementioned challenges and developments, we experimentally demonstrate the effectiveness of the unknown input decentralized mixed algebraic and dynamic state observation algorithm derived by part of the authors previously in \cite{lorenz-meyer_pmu-based_2020-1}. In this setting our contributions are three-fold:
\begin{itemize}
	\item In close cooperation with the German TSO 50Hertz Transmission GmbH and using a PMU provided by Studio Elektronike Rijeka d.o.o., we obtained PMU-measurements from the extra-high voltage grid of Germany at a substation close to a power plant. This measurement location is different from the terminal bus of the SG considered in \cite{lorenz-meyer_pmu-based_2020-1} and most of the publications in the literature, e.g., \cite{paul_dynamic_2018, ghahremani_local_2016, taha_risk_2018}. Thus, we map the recorded PMU-measurements to the terminal bus of the SG, making our algorithm applicable. 
	This mapping considers the tap-changing transformer and the auxiliary system of the power plant.
	\item  We relax the assumptions made in \cite{lorenz-meyer_pmu-based_2020-1}. More precisely, we remove the assumption on the direct-axis transient reactance $x'_\mrm{d}$ and the quadrature-axis reactance $x_\mrm{q}$ being equal and on the stator resistance being negligible. Also, we extend the algorithm to a case with time-varying and known mechanical torque $T_\mrm{m}$, e.g. modeled using a governor and turbine model. Furthermore, we present a novel layout scheme for designing the dynamic regressor extension and mixing (DREM)-filters, which shows to largely improve the convergence of the parameter identification using the recorded real-world PMU-measurements. Moreover, we develop an improved filtering with better attenuation of measurement noise and adaptive estimator gains. In the recorded real-world PMU-measurements this greatly simplified the application of the DREM-based parameter identification in different operational conditions without the need to re-tune the estimator gains.
	\item We utilize the recorded real-world PMU-measurements to experimentally validate our unknown input DSE algorithm. For this, we perform an auto- as well as a cross-validation. That is, we use two time-series from different operational regimes of the SG, of which only one is employed in tuning the algorithm, while the other one is used for the cross-validation.
\end{itemize}

The remainder of the paper is organized as follows. In Section~\ref{sec:problem_state}, the model of the considered power system is introduced. The mapping of the PMU-measurements is derived in Section~\ref{sec:mapping}. Extensions of the decentralized mixed algebraic and dynamic state observation from \cite{lorenz-meyer_pmu-based_2020-1} are presented in Section~\ref{sec:extensions}. By using real-world PMU-measurements, the proposed algorithm is experimentally validated in Section~\ref{sec:vali}. Finally, conclusive remarks and a brief outlook on future work are given in Section~\ref{sec:conclusions}.

\section{Real-world setup and employed SG model}
\label{sec:problem_state}
\subsection{Real-world setup}
\begin{figure}
	\centering
	\includegraphics[width=0.2\textwidth]{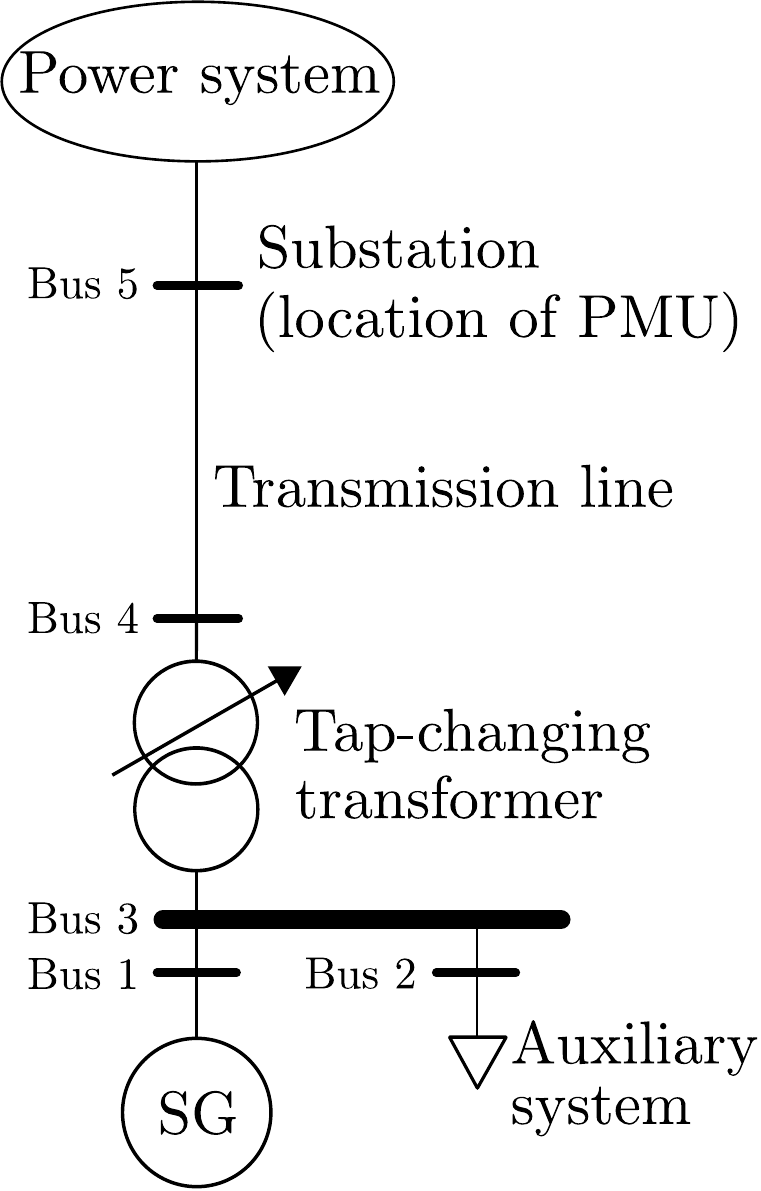}
	\caption{Schematic represenation of the considered power plant, including tap-changing transformer, auxiliary system and a PMU connected at the nearest substation (Bus 5).
	}
	\label{fig:Overview}
\end{figure}

For the work in the present paper, we consider a power plant in the 500 - 1000 MW class within the extra-high voltage grid (410~kV) of Germany. At a substation close to this power plant, a PMU is installed. 
This PMU positioning is advantageous and realistic as the substation is typically operated by the TSO. Thus, for an installation at this point no coordination with the plant operator is required. A typical connection of such a substation to a SG is shown in Figure \ref{fig:Overview}. The SG's terminal bus is denoted as Bus 1. The SG is connected to a common bus (Bus 3) together with the auxiliary system of the power plant, which is connected to Bus 2. Bus 3 is connected to the substation via a tap-changing transformer and a HV transmission line. 

In this setup, the employed SG model is given below. The models of the transmission line, the tap-changing transformer and the auxiliary system of the power plant are introduced in Section~\ref{sec:mapping} together with a mapping of the PMU-measurements taken at Bus 5 to the terminal bus of the SG (Bus 1)

\subsection{Model of the SG}
The SG dynamics at the power plant is represented by the well-known third-order flux-decay model, see e.g., \cite[Eq. (3.3), (3.5), (3.15)]{van_cutsem_voltage_1998},  \cite[Eq. (5.135-5.137)]{sauer_power_2006} and Model 1.0 described in Table 1 on page 17 in \cite{noauthor_ieee_2020}, i.e.,
\begin{subequations}
	\lab{model}
	\begali{
	\label{eq:x1}	\dot x_{1}&= x_{2}  ,\\
	\label{eq:x2}	\dot x_{2}&=\frac{\omega_\mrm{s}}{2H}(T_\mrm{m}-T_\mrm{e}-Dx_{2})  ,\\
		\dot x_{3} &= \frac{1}{T_\mrm{d0}'}(-x_{3}-(x_\mrm{d}-x_\mrm{d}')I_\mrm{td}+E_\mrm{f}) ,}
\end{subequations}
where we have defined the unknown state vector as
\begalis{
	x&\coloneqq\begmat{ x_{1} & x_{2} & x_{3} \end{bmatrix}^\top=\begin{bmatrix} \delta & \omega-\omega_\mrm{s} & E_\mrm{q}' }^\top,
}
with $\omega$ being the shaft speed, $\omega_\mrm{s}$ the nominal synchronous speed, $\omega_\mrm{t}$ the terminal voltage speed, $\delta$ the rotor angle, $E_\mrm{q}'$ the quadrature-axis internal voltage, $E_\mrm{f}$ the field voltage and $T_\mrm{e}$ the electrical air-gap torque and $V_\mrm{t}$ the terminal voltage magnitude. Moreover, the unknown constants are the inertia constant $H$, the damping factor $D$, the mechanical power $T_\mrm{m}$, the direct-axis transient reactance $x_\mrm{d}'$, the direct-axis reactance $x_\mrm{d}$  and the direct-axis transient open-circuit time constant $T_\mrm{d0}'$. 

The stator algebraic equation for the third-order model reads (see \cite[Eq. (5.134)]{sauer_power_2006})
\begin{equation}
	\begin{split}
		\label{eq:stator_alg}
\mrm{j} x_{3}\mrm{e}^{\mrm{j}(x_{1}-\frac{\pi}{2})} &= (R_\mrm{s} + \mrm{j} x_\mrm{d}') I_\mrm{t} \mrm{e}^{\mrm{j}\phi_\mrm{t}} +\\&+ V_\mrm{t} \mrm{e}^{\mrm{j}\theta_\mrm{t}} - (x_\mrm{q} - x_\mrm{d}') I_\mrm{tq}\mrm{e}^{\mrm{j}(x_{1}-\frac{\pi}{2})}, 
\end{split}
\end{equation}
where $R_\mrm{s}$ is the stator resistance, $x_\mrm{q}$ is the quadrature-axis reactance, $\theta_\mrm{t}$ is the terminal voltage angle and $\phi_\mrm{t}$ is the terminal current angle. Thus, the terminal current $I_\mrm{t}$ and active and reactive powers ($P_\mrm{t}$ and $Q_\mrm{t}$) can be expressed as
\begin{equation}
	\begin{split}
		I_\mrm{td} &= \frac{1}{R_\mrm{s}^2 + x_\mrm{d}' x_\mrm{q}}(x_\mrm{q}x_{3}  - x_\mrm{q} V_\mrm{q} + R_\mrm{s} V_\mrm{td}), \\
		I_\mrm{tq} &= \frac{1}{R_\mrm{s}^2 + x_\mrm{d}' x_\mrm{q}}(R_\mrm{s}x_{3}  - R_\mrm{s} V_\mrm{tq} + x_\mrm{d}' V_\mrm{td}), \\
		I_\mrm{t} &= \sqrt{I_\mrm{td}^2+I_\mrm{tq}^2},\\
		P_\mrm{t} &=\frac{1}{R_\mrm{s}^2 + x_\mrm{d}' x_\mrm{q}} (x_{3}(x_\mrm{q}V_\mrm{td}+R_\mrm{s}V_\mrm{tq}) +\\ &+V_\mrm{td}V_\mrm{tq}(x_\mrm{d}'-x_\mrm{q}) +R_\mrm{s}(V_\mrm{td}^2-V_\mrm{tq}^2)),\\
		Q_\mrm{t} &=\frac{1}{R_\mrm{s}^2 + x_\mrm{d}' x_\mrm{q}} (x_{3}(x_\mrm{q}V_\mrm{tq}-R_\mrm{s}V_\mrm{td}) +\\&+2R_\mrm{s}V_\mrm{td}V_\mrm{tq} -x_\mrm{d}'V_\mrm{td}^2- x_\mrm{q}V_\mrm{tq}^2 ).
	\end{split}
\end{equation}

We define the unknown constants
\begin{equation}
	\begin{split}
		a_{1}&=\frac{\omega_\mrm{s} D}{2H}, \ \ a_{2}= \frac{\omega_\mrm{s}}{2H}, 
	\end{split}
\end{equation}	
and can thus write the model \eqref{model} for the SG compactly as 
\begin{subequations}
	\lab{x}
	\begali{
		\lab{x1}
		\dot x_{1}&=x_{2},\\
		\lab{x2}	
		\dot x_{2}&= -a_{1} x_{2}+ a_{2}(T_\mrm{m} - T_\mrm{e}) ,\\
		\label{x3}
		\dot x_{3} &= \frac{1}{T_\mrm{d0}'}(-x_{3}-(x_\mrm{d}-x_\mrm{d}')I_\mrm{td}+E_\mrm{f}).
	}
\end{subequations}
\begin{figure}
	\centering
	\includegraphics[width=0.5\textwidth]{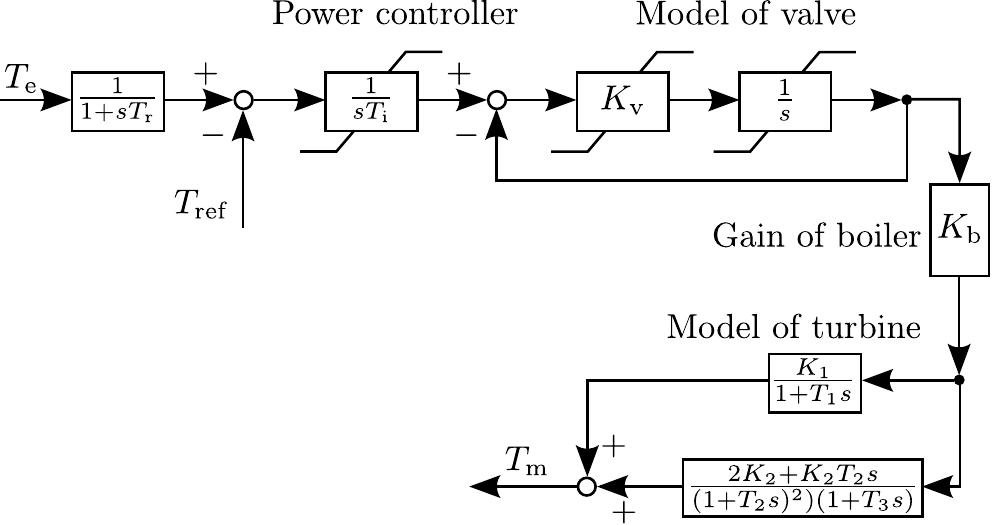}
	\caption{Model of the governor and turbine.}
	\label{fig:turbine_governor}
\end{figure}
The mechanical torque $T_\mrm{m}$ is assumed time-varying and modeled by a standard power control governor and turbine model used by 50Hertz Transmission GmbH, where the torque reference $T_\mrm{ref}$ is assumed constant and known. 
The governor and turbine model is shown in Figure \ref{fig:turbine_governor}, where $s$ denotes the Laplace operator. 

\section{Mapping PMU-measurements}
\label{sec:mapping}
In the following, we present the mathematical models needed to derive a map of the PMU-measurements from the substation to the terminal bus of the SG, i.e., Bus 1 in Figure~\ref{fig:Overview}. This mapping is essential as our algorithm, introduced in \cite{lorenz-meyer_pmu-based_2020-1} and most of the publications in the literature, e.g., \cite{paul_dynamic_2018, ghahremani_local_2016, taha_risk_2018}, assume PMU-measurements available at the terminal bus of the SG. Thus, by applying the mapping these algorithms can be utilized in the considered case of measuring at a substation close to a power plant. This is advantageous, as this measurement location is far more accessible to TSO's than the terminal bus of a SG. The corresponding mapping involves the HV transmission line, the tap-changing transformer and the auxiliary system of the power plant, for which there are - to the best of the author's knowledge - no standard models, as those highly depend on the individual setup of the power plant. \\
The PMU-measurement vector $\bm{y}^\mrm{PMU}$ is denoted by
\begin{equation}
	\bm{y}^\mrm{PMU}=\begin{bmatrix} V^\mrm{PMU}&  \theta^\mrm{PMU} & I^\mrm{PMU} & \phi^\mrm{PMU} \end{bmatrix}^\top ,
\end{equation}
 where $V^\mrm{PMU}$ is the PMU voltage magnitude, $\theta^\mrm{PMU}$ the PMU voltage angle, $I^\mrm{PMU}$ the PMU current magnitude and $\phi^\mrm{PMU}$ the PMU current angle.
 The mapped terminal bus measurement vector $\bm{y}$ is denoted by 
 \begin{equation}
 	\bm{y}=\begin{bmatrix} V_\mrm{t}&  \theta_\mrm{t} & I_\mrm{t} & \phi_\mrm{t} \end{bmatrix}^\top.
 \end{equation}
 
  We assume that the positive sequence components of the voltages and currents are significantly larger in magnitude than the negative and zero sequence components. Thus, before applying the mapping to the measurement data we decompose the latter in symmetrical components and consider only the positive sequence for further modeling and signal processing. 

\begin{remark}
	All variables are given in per unit values. 
\end{remark}

\subsection{Model of the HV transmission line}
\begin{figure}
	\centering
	\includegraphics[width=0.5\textwidth]{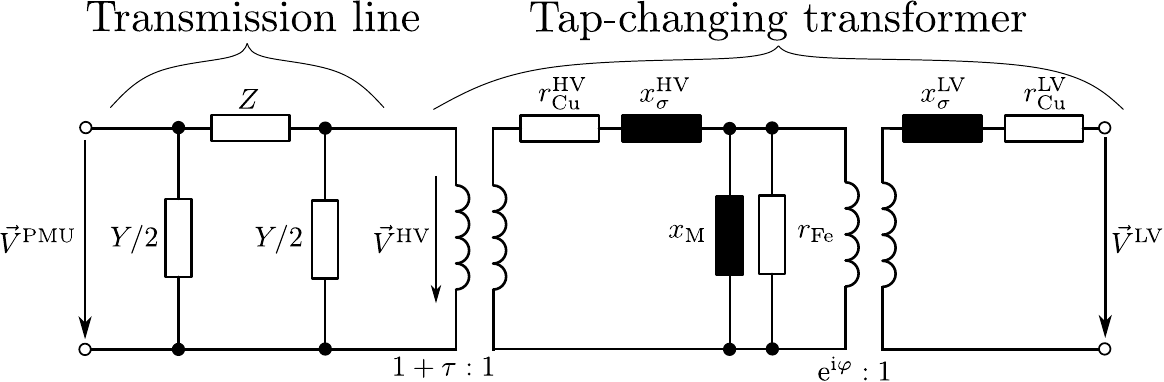}
	\caption{$\pi$-equivalent model of transmission line and tap-changing transformer.
	}
	\label{fig:Line_Trafo}
\end{figure}
The HV transmission line connecting Bus 4 with Bus 5 is modeled using a standard $\pi$-equivalent model \cite[Chap. 3.1]{machowski_power_2008}, shown in the left hand-side of Fig. \ref{fig:Line_Trafo}. The model maps the PMU voltage $\vec{V}^\mrm{PMU}$ and current $\vec{I}^\mrm{PMU}$ phasors recorded at the substation (Bus 5) to the HV side of the tap-changing transformer ($\vec{V}^\mrm{HV}$ and $\vec{I}^\mrm{HV}$ at Bus 4) and is given as
\begin{equation}
	\label{eq:line}
	\begin{split}
		\vec{V}^\mrm{HV} &= (1 + \frac{1}{2} ZY) \vec{V}^\mrm{PMU} - Z \vec{I}^\mrm{PMU}   ,\\
		\vec{I}^\mrm{HV} &= (Y+\frac{ZY^2}{4})\vec{V}^\mrm{PMU} -(1 + \frac{1}{2} ZY) \vec{I}^\mrm{PMU}  , 
	\end{split}
\end{equation}
where $Z$ represents the total series impedance and $Y$ the total shunt admittance of the line. 

\subsection{Model of the tap-changing transformer}
The transformer is represented using a standard T-equivalent model \cite[Chap. 3.2]{glover_power_2012}, shown in the right hand-side of Figure \ref{fig:Line_Trafo}. The tap-changer is modeled by an additional, ideal transformer connected to the high-voltage (HV) side. The model maps the voltage and current phasors at the HV side of the tap-changing transformer ($\vec{V}^\mrm{HV}$ and $\vec{I}^\mrm{HV}$) to the low-voltage (LV) side ($\vec{V}^\mrm{LV}$ and $\vec{I}^\mrm{LV}$), i.e., to Bus 3, and is given as
\begin{equation}
	\begin{split}
		\label{eq:trafo}
		\vec{I}^\mrm{LV} &= \mrm{e}^{-\mrm{j}\varphi} \Biggl(\frac{\vec{V}^\mrm{HV}(r_\mrm{Fe} +\mrm{j}x_\mrm{M})}{(1+\tau) \mrm{j}r_\mrm{Fe} x_\mrm{M}} +\vec{I}^\mrm{HV}(1+\tau)+  \\
		& +\vec{I}^\mrm{HV}(1+\tau)\frac{(r^\mrm{HV}_\mrm{Cu} +\mrm{j}x^\mrm{HV}_\mrm{\sigma})(r_\mrm{Fe} +\mrm{j}x_\mrm{M})}{\mrm{j}r_\mrm{Fe} x_\mrm{M}}\Biggr)   , \\
		\vec{V}^\mrm{LV} &=\vec{I}^\mrm{LV} (r^\mrm{LV}_\mrm{Cu} +\mrm{j}x^\mrm{LV}_\mrm{\sigma})+ \\&+ \mrm{e}^{-\mrm{j}\varphi}\left(\frac{\vec{V}^\mrm{HV}}{1+\tau} + \vec{I}^\mrm{HV}(r^\mrm{HV}_\mrm{Cu} +\mrm{j}x^\mrm{HV}_\mrm{\sigma})(1+\tau)\right)   , \\
	\end{split}
\end{equation}
where $r^\mrm{HV}_\mrm{Cu}$ and $r^\mrm{LV}_\mrm{Cu}$ are the winding resistance on the HV and LV sides, $x^\mrm{HV}_\mrm{\sigma}$ and $x^\mrm{LV}_\mrm{\sigma}$ are the leakage reactances on the HV and LV sides, $x_\mrm{M}$ is the magnetizing reactance and $r_\mrm{Fe}$ is the parallel shunt resistance. The phase shift of the transformer is denoted by $\varphi$ and the additional voltage per tap is denoted by $\tau$.

\subsection{Model of the auxiliary system of the power plant}
Detailed modeling of the auxiliary system of a power plant depends to a large extent on the specific setup of the individual power plant and is, in general, not completely known to the TSO. Nonetheless, it needs to be considered in the overall power balance between the total power generated by the plant and the power measured by the PMU at the substation. Thus, for the purposes of the present paper we propose to model the active power demand of the auxiliary system in dependency of the average power generation of the power plant over a time-window $[t_{0}, t_1]$, $t_1>t_0.$ The reactive power demand is modeled with a constant power factor to account for the induction machines typically present in the auxiliary system \cite[Chap. 7]{oeding_elektrische_2011}. 	\begin{equation}
	\begin{split}
		\label{eq:PQ_AS}
		P^\mrm{AS} &=\frac{P^\mrm{AS}_{\text{max}}} {P^\mrm{SG}_{\text{max}}-P^\mrm{AS}_{\text{max}}}  \frac{1}{t-t_{0}}\int_{t_{0}}^{t}P^\mrm{LV}  \text{d}\rho   ,\\
		Q^\mrm{AS} &= P^\mrm{AS} \tan (\arccos(pf))   ,
	\end{split}
\end{equation}
where $P^\mrm{AS}$ and $Q^\mrm{AS}$ are the active and reactive power demand of the auxiliary system, i.e., at Bus 2, $P^\mrm{LV}$ is the active power at the LV side of the transformer, i.e., at Bus 3. $P^\mrm{SG}_{\text{max}}$ is the maximum power generation of the SG, $P^\mrm{AS}_{\text{max}}$ is the maximum power consumption of the auxiliary system and $pf$ is the power factor. 

Thus, the current phasor of the auxiliary system $\vec{I}^\mrm{AS} = I_{x}^\mrm{AS} +\mrm{j}I_{y}^\mrm{AS}$ can be derived from the calculated apparent power $\vec{S}^\mrm{AS} = P^\mrm{AS} + \mrm{j}Q^\mrm{AS}$ and the mapped LV side voltage phasor $\vec{V}^\mrm{LV} = V_{x}^\mrm{LV}+\mrm{j}V_{y}^\mrm{LV}$ as follows
\begin{equation}
	\begin{split}
	\label{eq:I_AS}
	I^\mrm{AS}_{x} &=\frac{1}{V^\mrm{LV}_{x}}\left(P^\mrm{AS} - V^\mrm{LV}_{y} I^\mrm{AS}_{y}\right)   , \\
	I^\mrm{AS}_{y} &=\frac{1}{\left( \left(V^\mrm{LV}_{x}\right)^2+\left(V^\mrm{LV}_{y}\right)^2\right)} \left(P^\mrm{AS}V^\mrm{LV}_{y} - Q^\mrm{AS}V^\mrm{LV}_{x}\right)   .
\end{split}
\end{equation}
The current and voltage phasors at the SG terminal (Bus 1) can be expressed as 
\begin{equation}
	\begin{split}
		\vec{V}^\mrm{SG} &=\vec{V}^\mrm{LV}   ,\\
		\vec{I}^\mrm{SG} &=\vec{I}^\mrm{LV}+\vec{I}^\mrm{AS}  . 
	\end{split}
	\label{mapping}
\end{equation}
\subsection{PMU-measurements mapped to SG terminal}
By sequentially applying \eqref{eq:line}-\eqref{mapping} to the PMU-measurements $\bm{y}^\mrm{PMU}$, we obtain the mapping to the SG terminal $$\bm{y}^\mrm{PMU} \mapsto \bm{y}.$$ In the sequel, we assume the mapping was already performed and denote the mapped measurements at the terminal of the SG as
\begin{equation}
	\begin{split}
		\label{eq:y}
		\bm{y} &=\begin{bmatrix}y_{1} & y_{2} &y_{3} & y_{4}  \end{bmatrix} \\&= \begin{bmatrix} |\vec{V}^\mrm{SG}| &  \arg \{\vec{V}^\mrm{SG}\} &|\vec{I}^\mrm{SG}| &  \arg \{\vec{I}^\mrm{SG}\}  \end{bmatrix} \\&= \begin{bmatrix} V_\mrm{t} &  \theta_\mrm{t} &I_\mrm{t} &  \phi_\mrm{t}  \end{bmatrix}    ,
	\end{split}
\end{equation}
where $|\cdot|$ denotes the magnitude and  $\arg\{\cdot\}$ the argument of a complex number. The mapped measurements can then be fed to the decentralized mixed algebraic and dynamic state observer derived in \cite{lorenz-meyer_pmu-based_2020-1}, see Figure~\ref{fig:obs_struct}.

\section{Practical extensions of the decentralized mixed algebraic and dynamic state observer}
\label{sec:extensions}
We present several extensions of the decentralized mixed algebraic and dynamic state observation algorithm introduced in~\cite{lorenz-meyer_pmu-based_2020-1} (see also Figure \ref{fig:obs_struct}), which are instrumental for its successful practical implementation in Section~\ref{sec:vali}. 

More precisely, in Section~\ref{subsec:algob} we present a new version of the algebraic observer for $x_{1}$ and $x_{3}$ in \eqref{x} under relaxed assumptions, while in Section~\ref{Sec:modified_parameter_estimator} we propose several extensions of the DREM-based I\&I adaptive observer for $x_{2}$ in \eqref{x}.
In summary, compared to the algorithm presented in in~\cite{lorenz-meyer_pmu-based_2020-1}, these extensions and modifications allow us to significantly ease the required assumptions and substantially improve the algorithm's performance when using real-world PMU-measurements. At the same time, they support a structured implementation and tuning procedure, which we hope eases the accessibility of our proposed approach for end-users, e.g., in the industry.

\begin{figure}
	\centering
	\includegraphics[width=0.5\textwidth]{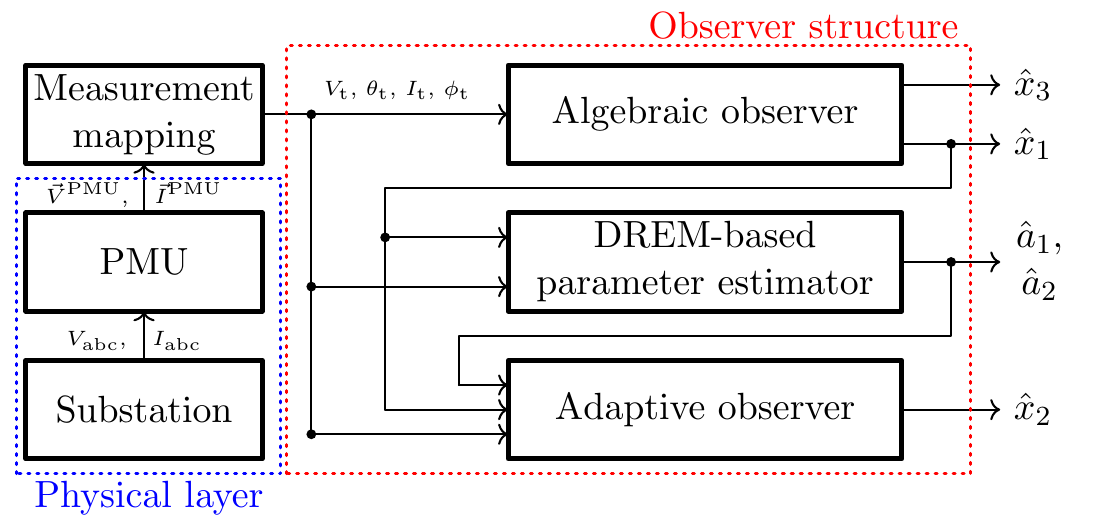}
	\caption{Structure of the proposed algebraic observer in combination with the DREM-based I\&I adaptive observer including the mapping of the PMU-measurements.}
	\label{fig:obs_struct}
\end{figure}

\subsection{Extension of the algebraic observer for $x_{1}$ and $x_{3}$}
\label{subsec:algob}
In \cite{lorenz-meyer_pmu-based_2020-1}, it was assumed for the SG model \eqref{x}, \eqref{eq:stator_alg} that the direct-axis transient reactance $x'_\mrm{d}$ and the quadrature-axis reactance $x_\mrm{q}$ are equal and that the stator resistance is negligible. We relax both of these assumptions, since they may be difficult to verify in applications. 

Some algebraic operations on the stator algebraic equation~\eqref{eq:stator_alg} allow us to explicitly compute the unmeasurable states $x_{1}$ and $x_{3}$, requiring only the knowledge of the quadrature-axis reactance $x_\mrm{q}$, the direct-axis transient reactance $x_\mrm{d}'$ and the stator resistance $R_\mrm{s}$. This observation is summarized in the lemma below, whose proof is given in Appendix \ref{app:proofs}. 
\begin{lemma}\em
	\lab{pro1}
	The states $x_{1}$ and $x_{3}$ of the system \eqref{x}, \eqref{eq:stator_alg} can be determined uniquely from the mapped PMU-measurements \eqref{eq:y} via 
	\begin{subequations}
		\lab{obs}
		\begin{align}
			\label{eq:alg_x1} x_{1} &= \arg\{\vec{\psi}\}   ,\\
			\label{eq:alg_x3} x_{3} &= |\vec{\psi}| - (x_\mrm{q}-x'_\mrm{d})\cos\left(\frac{\pi}{2}-x_{1}+y_{4}\right)y_{3}  ,
		\end{align}
	\end{subequations}
	where the phasor $\vec{\psi}$ corresponds to 
	\begin{equation*}
		  \vec{\psi} \coloneqq (R_\mrm{s} + \mrm{j} x_\mrm{q}) y_{3} \mrm{e}^{\mrm{j}y_{4}} + y_{1} \mrm{e}^{\mrm{j}y_{2}}.
	\end{equation*}
\end{lemma}

With $x_3$ known from \eqref{eq:alg_x3}, the electrical air-gap torque $T_\mrm{e}$ can be expressed as
\begin{equation}
	T_\mrm{e} = (x_\mrm{q}-x_\mrm{d}')I_\mrm{d}I_\mrm{q}+x_3I_\mrm{d}.
\end{equation}
\begin{remark}
	A related result is given in \cite{uecker_differential_2015}, where it is shown that the so-called single machine infinite bus system, with the SG modeled by the flux-decay model \eqref{x}, is a differentially flat system. However, in \cite{uecker_differential_2015} this property is used for trajectory planning, open-loop control and a simple linearizing feedback scheme, but not for state observation. 
\end{remark}
\subsection{Practical extensions of the DREM-based I\&I adaptive observer for $x_{2}$}
Building on the DREM-based I\&I adaptive observer for $x_{2}$ introduced in \cite[Sec. 4]{lorenz-meyer_pmu-based_2020-1}, we propose the following extensions, which in our experience proved to significantly improve the algorithms performance when using real-world PMU-measurements.
\begin{enumerate}
	\item An extension of the parameter estimator to the case of time-varying and known mechanical torque $T_\mrm{m}$.
	\item A regressor extension based on approximating its time-derivatives. This regressor extension is inspired by the linear independence test using the Wronskian (see \cite[Sec. 18.4.4]{poznyak_advanced_2008}) and showed significantly improved convergence and decreased excitation requirements using the recorded real-world PMU-measurements in comparison to the one based on delay operators proposed in \cite[Eq. 20]{lorenz-meyer_pmu-based_2020-1}.
	\item An improved filtering design which shows better attenuation of measurement noise when using the recorded real-world PMU-measurements. Furthermore, adaptive estimator gains, which improve the parameter estimators performance in presence of time-varying excitation levels in the recorded real-world PMU-measurements. 
	\item A simplified observer structure utilizing the measured voltage and current phase angles ($\theta_\mrm{t}$ and $\phi_\mrm{t}$) instead of the voltage frequency.
\end{enumerate}
In the following, we give more details about each proposed extension.
\subsubsection{Practical modifications of the parameter estimator}
\label{Sec:modified_parameter_estimator}
We extend the DREM-based parameter estimator to the case of time-varying and known mechanical torque $T_\mrm{m}$. This is possible if, for example, a model for the governor and turbine is available and used to generate $T_\mrm{m}$. The case of unknown and constant mechanical torque $T_\mrm{m}$ is addressed in \cite{lorenz-meyer_pmu-based_2020-1}. 
Thus, the vector of unknown parameters can be defined as follows 
\begequ
\begin{split}
	\lab{the}
	\bm{\theta} \coloneqq \begin{bmatrix} a_{1}&a_{2}	\end{bmatrix}^\top. \\
\end{split}
\endequ 
Details of the derivation can be found in Appendix \ref{app:para_est}.
\subsubsection{Practical modifications of the regressor extension}
\label{sec:mod_regressor_ext}
The suitable design of the regressor extension is a fundamental, yet often non-trivial, step in implementing the DREM approach~\cite{aranovskiy_performance_2017}. 
By taking inspiration from the linear independence test using the Wronskian (see \cite[Sec. 18.4.4]{poznyak_advanced_2008}), we propose to extend the regressor with the following linear, bounded-input bounded-output (BIBO)-stable operator $\bm{\mathcal{H}}$ 
\begin{equation}
	\begin{split}
		\bm{\mathcal{H}}&= K\begin{bmatrix}  1 & \frac{ c_1 c_2 s}{(c_1+s) (c_2 + s)}\end{bmatrix}^\top ,\\
	\end{split}
\end{equation}
with positive real gain $K$ and tuning parameters $c_j>0, \ j={1,2}$. The second output of the operator $\bm{\mathcal{H}}$ approximates the first-time derivative of its input. In this way, the second output is phase shifted w.r.t. the first output by approximately 90 degrees within a tuneable bandwidth. This is advantageous as the goal of extending the regressor $\psi$ (see \eqref{eq:regressor_ext} in Appendix \ref{app:para_est}) is to generate a square regressor $\Psi$ with linearly independent rows. This is facilitated by the introduced phase shift.
\subsubsection{Improved filtering and adaptive estimator gains}
\label{sec:improved_filtering}
Instead of the second-order delay filter proposed in \cite[Eq. 9]{lorenz-meyer_pmu-based_2020-1}, we utilize the following third-order delay filter 
	\begin{equation}
	\mathcal{F} = \frac{\lambda_1\lambda_2\lambda_3}{(\lambda_1+s)(\lambda_2+s)(\lambda_3+s)},
\end{equation}
with tuning parameters $\lambda_j>0, \ j={1,2,3}$. This filter attenuates high frequency signal components and thus decreases the influence of measurement noise. Furthermore, we scale the parameter estimator gains in dependence of the amount of excitation present over a moving average window by an additional time-varying gain $K^\gamma_{j} \in \ [\epsilon, 100]$, $j = 1,2$, where $\epsilon>0$ denotes the lower bound of the additional gain. 
Hence, during periods of low excitation the estimator gain is increased and vice versa (see \eqref{eq:para_est} and \eqref{eq:Kj} in Appendix \ref{app:para_est}).
\subsubsection{Simplified adaptive I\&I Observer}
\lab{subsec33}
By utilizing the measured voltage and current phase angles, i.e., $\theta_\mrm{t}$ and $\phi_\mrm{t}$ in \eqref{eq:y}, the DREM-based I\&I adaptive observer introduced in \cite[Eq. 17]{lorenz-meyer_pmu-based_2020-1} can be simplified as stated in the Lemma below.

\begin{lemma}\em
	\lab{pro2}
	Consider the dynamics \eqref{eq:x2} with $x_{1}$ from \eqref{eq:alg_x1}. 
	Define the DREM-based I\&I adaptive observer as
\begin{equation}
	\begin{split} 
		\dot{x}^{I}_{2} &= - ({\hat \theta_{1}}+k) ({x}^{I}_{2} + k x_{1}) +\hat \theta_{2} (T_\mrm{m}-T_\mrm{e}) ,\\
		\hat{x}_{2} &= {x}^{I}_{2} + k x_{1},
	\end{split} 
	\label{eq:obsx2}
\end{equation}
where $k>0$ is a tuning parameter. 
Then, if the excitation assumption \eqref{eq:exccon} is satisfied,
$$
\lim_{t\to\infty}\tilde {x}_{2}(t)=0,\quad \lim_{t\to\infty}\tilde\theta=0,\quad i\in\{1,2\},
$$
where $\tilde{x}_2=\hat{x}_2-x_2$ is the state observation error and $\tilde\theta=\hat\theta-\theta$ is the parameter estimation error.
\end{lemma}
 The proof follows the same procedure as depicted in \cite[Sec. 4]{lorenz-meyer_pmu-based_2020-1} and is not included here due to the space limitations. 
Provided the excitation assumption \eqref{eq:exccon} is fulfilled, convergence of the adaptive observer is guaranteed via cascaded systems stability analysis (see e.g., \cite{vidyasagar_decomposition_1980}).

\section{Experimental validation}
\label{sec:vali}
In this section, we present the experimental validation of the proposed algorithm, see Figure~\ref{fig:obs_struct}, using real-world PMU-measurements\footnote{
For detailed simulation results employing the New England IEEE 39 bus system \cite{ramos_et_al_benchmark_2015}, the reader is referred to \cite{lorenz-meyer_pmu-based_2020-1}.}. The measurements were acquired in cooperation with the German TSO 50Hertz Transmission GmbH at a substation in Germany close to a power plant as depicted in Figure \ref{fig:Overview}. The PMU was provided by Studio Elektronike Rijeka d.o.o.\footnote{For further information, see http://www.ster.hr/}.

To validate the proposed algorithm given in Figure~\ref{fig:obs_struct}, at first we map the acquired PMU-measurements to the terminal bus of the SG (Bus 1 in Figure~\ref{fig:Overview}) by using    \eqref{eq:line}-\eqref{mapping}. Subsequently, we estimate $x_{1}$ and $x_{3}$ algebraically via Lemma~\ref{pro1}. Then, we follow a two-step approach. 
\begin{itemize}
	\item Step 1: We employ the adaptive observer of Section \ref{Sec:modified_parameter_estimator} to estimate the state $x_{2}$ and the parameters $a_{1}, a_{2}$. As the reference for $a_{2}$ is known, the estimated value $\hat{a}_{2}$ is compared to the known reference $a_{2}$. Furthermore, the estimated state $\hat{x}_{2}$ is compared to the mapped PMU-measurement $x_{2}$. 
	\item Step 2: To verify the estimated parameters we simulate the SG model according to \eqref{x}, using the mapped measurement of the terminal voltage magnitude $V_\mrm{t}$ and angle $\theta_\mrm{t}$ as well as an estimate of the exciter voltage $E_\mrm{f}$ obtained via
	$$ \hat{E}_\mrm{f} = T_\mrm{d0}'\dot{\hat{x}}_3 +\hat{x}_3 +(x_\mrm{d}-x_\mrm{d}')I_\mrm{td}, $$
	where $\hat{x}_3$ is estimated according to \eqref{eq:alg_x3} and its time derivative $\dot{\hat{x}}_3$ is numerically calculated. We carry out this simulation using the parameters obtained via the DREM-based I\&I adaptive observer. The simulation results, more specifically the simulated $x^{\text{sim}}_{2}$, $I^{\text{sim}}_\mrm{t}$, $P^{\text{sim}}_\mrm{t}$, $Q^{\text{sim}}_\mrm{t}$, are compared to the mapped PMU-measurements. This step is commonly used for model validation in commercial software and is often referred to as "event playback" \cite{zhao_et_al_power_2021}.
\end{itemize}
This two-step approach is presented for a first measurement time-series, which was used to tune the parameters of the adaptive observer, i.e., the auto-validation. 

After this procedure is completed successfully, we employ a second time-series, which was not involved in tuning the parameters of the adaptive observer and where the SG operates at a much lower power setpoint, to perform the cross-validation. The latter comprises repeating Steps 1 and 2 with the tuning parameters obtained from the auto-validation.

To quantify the results, we use the symmetric mean absolute percentage error (sMAPE) as introduced below \cite{armstrong_long-range_1985}
\begin{equation}
	\text{sMAPE} = \frac{100 \%}{M} \sum_{k=1}^{M} \frac{|\hat{z}-z|}{0.5(|\hat{z}_{i}|+|z_{i}|)} ,
	\label{smape}
\end{equation}
where $z$ is the measured value and $\hat{z}$ is the estimated value or simulated value of the variable to be quantified, e.g., $\hat{x}_{2}$, $\hat{x}^{\text{sim}}_{2}$, $I^{\text{sim}}_\mrm{t}$, $P^{\text{sim}}_\mrm{t}$ or $Q^{\text{sim}}_\mrm{t}$. $M$ denotes the number of data points to be considered.\\
The torque reference $T_\mrm{ref}$, according to Figure~\ref{fig:turbine_governor}, is calculated from the experimental data assuming quasi steady-state conditions. The design parameters for the DREM-based I\&I adaptive observer are shown in Table \ref{tab:params}.
\begin{table}
	\centering
	\begin{tabular}{l|l|l}
		\cmidrule{1-3}
		Symbol & Description& Value  \\  	\cmidrule{1-3}
		$\lambda_1$ & Filter parameter & 8\\
		$\lambda_2$ & Filter parameter & 6.2\\
		$\lambda_3$ & Filter parameter & 7.4 \\
		$c_1$ & Filter parameter & 8\\
		$c_2$ & Filter parameter & 6\\
		$c_3$ & Filter parameter & 7 \\
		$K$ & Filter gain & 6.5 \\
		$\gamma_{1,2}$ &Adaptation gain (DREM) & $850$ \\
		$k$ & Observer gain& 8\\
		\cmidrule{1-3}
	\end{tabular}
	\caption{Employed design parameters for the DREM-based I\&I adaptive observer. }
	\label{tab:params}
\end{table} 
\begin{remark}
A direct validation of the results is infeasible in the considered scenario as the rotor angle $x_{1}$ and the virtual quantity of the quadrature-axis internal voltage $x_{3}$ can not be measured by a PMU located at a substation close to the SG. Hence, no reference values are known for these two states.
The performance of the algebraic observer according to Lemma~\ref{pro1} can thus only be validated indirectly through the DREM-based I\&I adaptive observer for $x_{2}$, as the performance of the observer critically relies on the knowledge of the correct $\hat{x}_{1}$ (see Lemma~\ref{pro2}). Furthermore, the parameter $a_{1}$ is unknown as it contains the damping factor $D$, which is introduced in the SG model to account for the simplifications made, when reducing the SG model order to four or lower, see \cite{sauer_power_2006}. 
\end{remark}
\begin{remark}
	The relative shaft speed of the rotor $x_{2}$ can not be measured directly by the PMU. As can be seen from \cite[Fig. 1]{lorenz-meyer_pmu-based_2020-1}, $x_{2}$ can be calculated from the mapped frequency of the voltage at the SGs terminal $f_\mrm{t}$ and the time derivative of the rotor angle, i.e.,
	\begin{equation}
		x_{2} = (2\pi f_\mrm{t} - \omega_\mrm{s}) + \dot{x}_{1}.
	\end{equation}
	by using the experimental measurements, $\dot{x}_{1}$ was numerically calculated employing $x_{1}$ estimated by the algebraic observer according to Lemma~\ref{pro1} and was found to be approximately two orders of magnitude smaller than $(2\pi f_\mrm{t} - \omega_\mrm{s})$.	Thus, for the experimental validation the influence of $\dot{x}_{1}$ on $x_{2}$ was neglected.
\end{remark}

\subsection{Auto-validation} 
We perform an auto-validation of the proposed algorithm following the two-step approach described previously. 
\begin{itemize}
	\item Step 1: For the first step, the results of the adaptive observer are shown in Figure \ref{fig:Autovali_Drem}. The observers' performance is evaluated using the sMAPE, see \eqref{smape}, to quantify the difference between the measured and estimated value of the state $x_{2}$. For the calculation of the sMAPE merely the values of $x_2$ after convergence of the parameter estimator are considered, i.e., after $t\approx88$~s. The calculated sMAPE is depicted in Table \ref{tab:sMAPE_auto}. It can be seen that the sMAPE has a very low value of below $0.4$~\% and the observer for $x_2$ hence shows a very good performance. 

For the parameter $a_2$ - for which a reference value is known - the parameter estimator shows a good result. The reference value is $a_2 = 25.41$, while the estimated values is $\hat{a}_2 = 26.65$. Thus, the estimate is very accurate with an error between the reference and the estimated value of only $5$~\%. 
\item Step 2: To further investigate the accuracy of the estimated parameters, we perform the second step of the validation procedure. For this, we simulate the mechanical part of the third-order SG model \eqref{x} using the identified parameters of the adaptive observer. The errors between the simulated and the mapped measurements, defined as 
\begin{equation}
	\label{eq:err}
	\begin{split}
		\tilde{x}_{2,\text{sim}} &= x_{2,\text{sim}} - x_{2} \ , \quad \tilde{I}_{t,\text{sim}} = I_{t,\text{sim}}- I_\mrm{t}, \\ \tilde{P}_{t,\text{sim}} &= P_{t,\text{sim}}- P_\mrm{t} \ , \quad \tilde{Q}_{t,\text{sim}} = Q_{t,\text{sim}} -Q_\mrm{t},
	\end{split}
\end{equation}
are shown in Figure \ref{fig:Autovali_sim} and the calculated sMAPE is depicted in Table \ref{tab:sMAPE_auto}. It can be seen that all sMAPE values are below $1.65$~\%. Thus, together with the accurately identified parameter $\hat{a}_2$ it can be concluded that for the auto-validation the adaptive observer shows very good results and is able to accurately estimate the unknown parameters as well as reconstruct the state $x_2$. 
\end{itemize}
\begin{figure}
	\centering
	\includegraphics[width=0.5\textwidth]{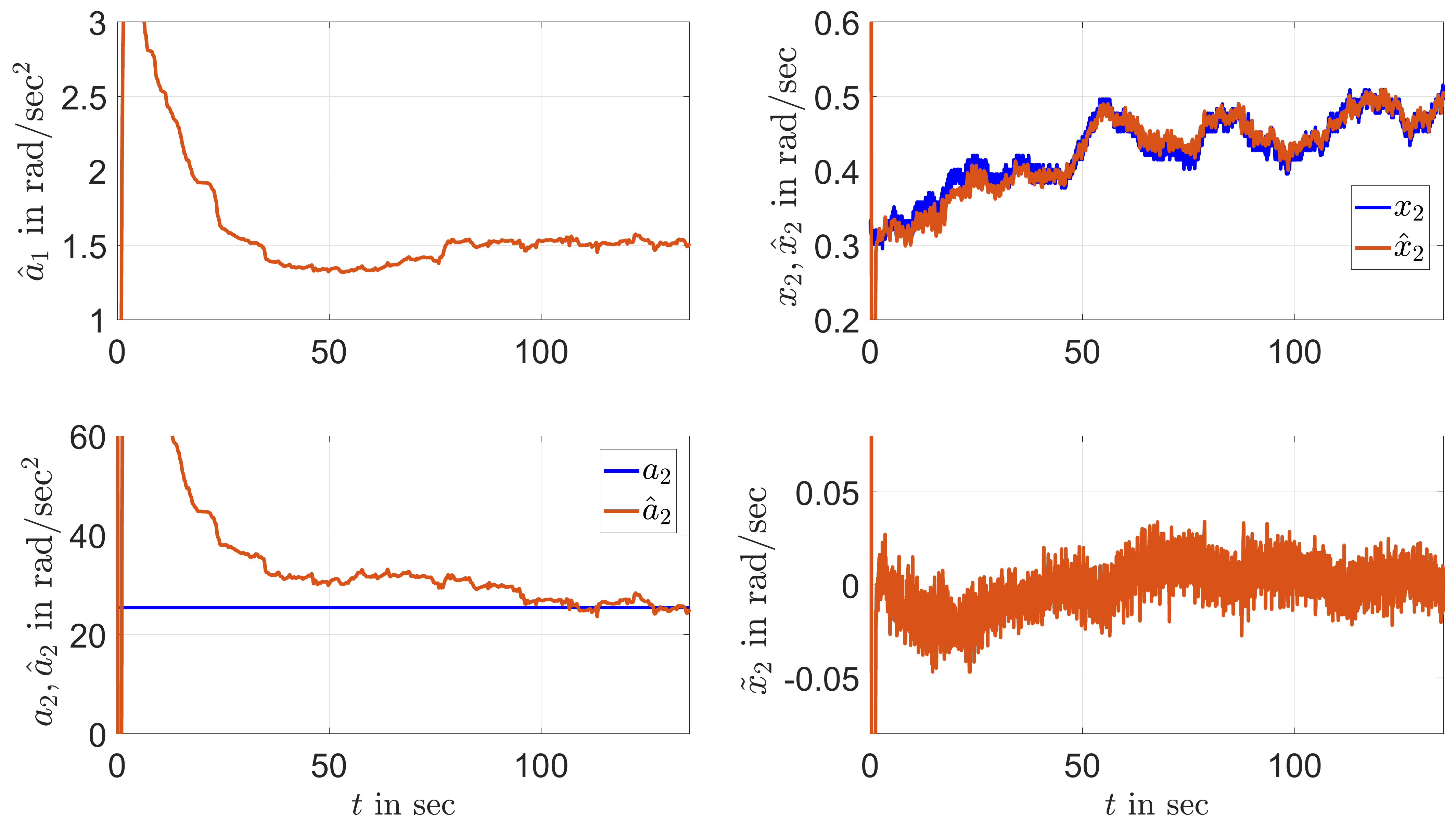}
	\caption{Auto-validation of the I\&I adaptive observer for $x_{2}$ and the DREM-based parameter estimation.}
	\label{fig:Autovali_Drem}
\end{figure}
\begin{figure}
	\centering
	\includegraphics[width=0.5\textwidth]{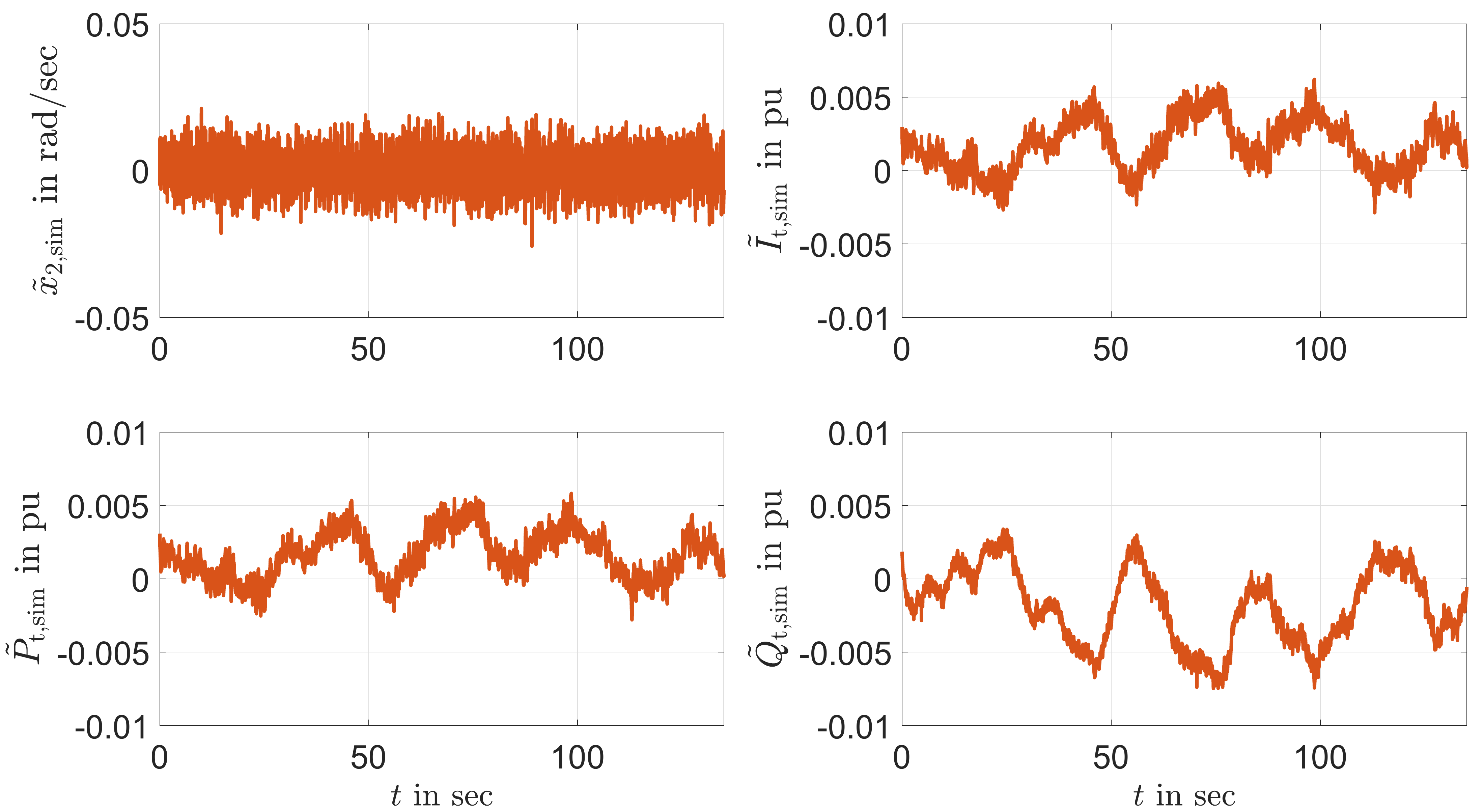}
	\caption{Error between the simulation results using the parameters obtained via the DREM-based parameter estimation and the mapped PMU-measurements for the auto-validation.}
	\label{fig:Autovali_sim}
\end{figure}
\begin{table}
	\centering
	\begin{tabular}{l|l|l}
	\cmidrule{2-3}
	& State &sMAPE in \% \\ \cmidrule{1-3}
	Adaptive observer  & $x_{2}$ & 0.37~\% \\ \cmidrule{1-3}
	&   $x_{2}$ & 0.24~\% \\
	Simulation	 & $I_\mrm{t}$ & 0.07~\% \\	
	results	&  $P_\mrm{t}$& 0.07~\% \\	
	&  $Q_\mrm{t}$ & 1.65~\% \\	\cmidrule{1-3}
	\end{tabular}
	\caption{sMAPE of the adaptive observer and the simulation results for the auto-validation.}
	\label{tab:sMAPE_auto}
\end{table}
\subsection{Cross-validation}
Next, we perform a cross-validation with experimental measurements, which were not employed for tuning the adaptive observer. The results are presented analogously to the auto-validation. \begin{itemize}
	\item Step 1: Figure \ref{fig:Crossvali_Drem} shows the results of the adaptive observer. The evaluated observer performance using the sMAPE is depicted in Table \ref{tab:sMAPE_cross}. For the calculation of the sMAPE merely the values of $x_2$ after convergence of the parameter estimator are considered, thus after $t\approx105$~s. It can be seen that the parameter estimator takes approximately $17$~s longer to converge and the sMAPE has a slightly increased value of about $1.2$~\%. The estimated value for the parameter $a_2$ is $\hat{a}_2 = 27.26$. Thus, there is an error between the reference and the estimated value of $7$~\%.
	\item Step 2: The error between the simulated values using the identified parameters and the mapped measurements (see \eqref{eq:err}), are shown in Figure \ref{fig:Crossvali_sim}. The calculated sMAPE is depicted in Table~\ref{tab:sMAPE_cross}. It can be seen that the the sMAPE values are again very low with a maximum value of $0.73$~\%. 
\end{itemize}
Therefore, we can conclude that overall the performance of the adaptive observer is very good. As to be expected, the results of the auto-validation are overall slightly better in relation to the cross-validation. 
The reconstruction of the state $x_2$ as well as the parameters works reliably for both time-series, thus for the auto- as well as the cross-validation. 

During the analysis of the real-world PMU-measurements the practical extensions introduced in Section \ref{sec:extensions} showed to significantly improve the algorithms performance. More precisely, the practical modification of the parameter estimator to the case of time-varying and known mechanical torque $T_\mrm{m}$ (see Section \ref{Sec:modified_parameter_estimator}) yields more accurate parameter estimates compared to the case of unknown and constant mechanical torque $T_\mrm{m}$ introduced in \cite{lorenz-meyer_pmu-based_2020-1}. This extension is feasible, as the standard power control governor and turbine model used by 50Hertz Transmission GmbH shown in Figure \ref{fig:turbine_governor} is available and was used to generate $T_\mrm{m}$. 
Employing the practical modifications of the regressor extension (see \ref{sec:mod_regressor_ext}), improved filtering and adaptive estimator gain (see \ref{sec:improved_filtering}) it was possible to successfully apply the algorithm to very different recorded PMU-measurement time-series with largely varying operation points utilizing the same design parameters. 
\begin{figure}
	\centering
	\includegraphics[width=0.5\textwidth]{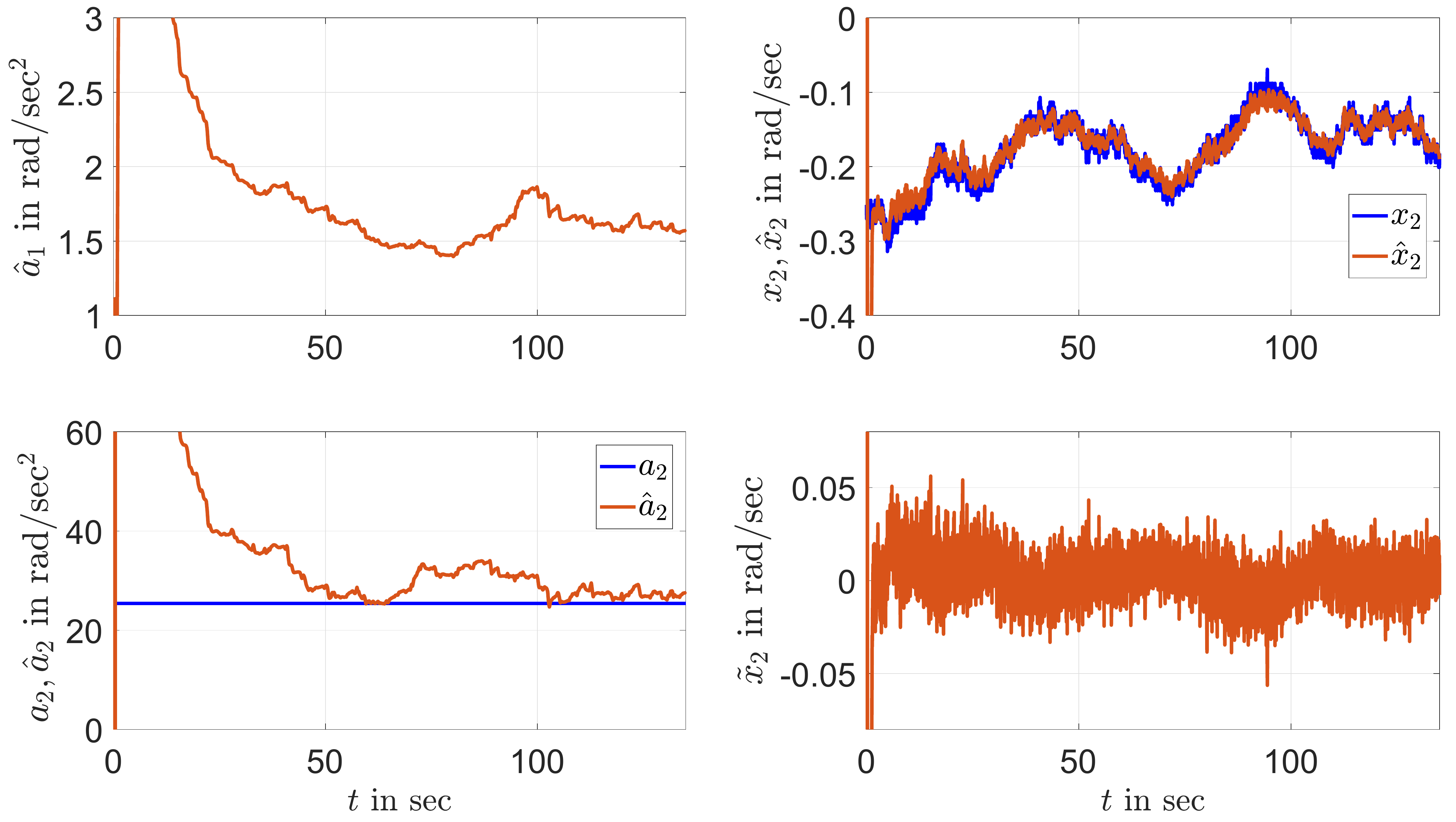}
	\caption{Cross-validation of the I\&I adaptive observer for $x_{2}$ and the DREM-based parameter estimation.}
	\label{fig:Crossvali_Drem}
\end{figure}
\begin{figure}
	\centering
	\includegraphics[width=0.5\textwidth]{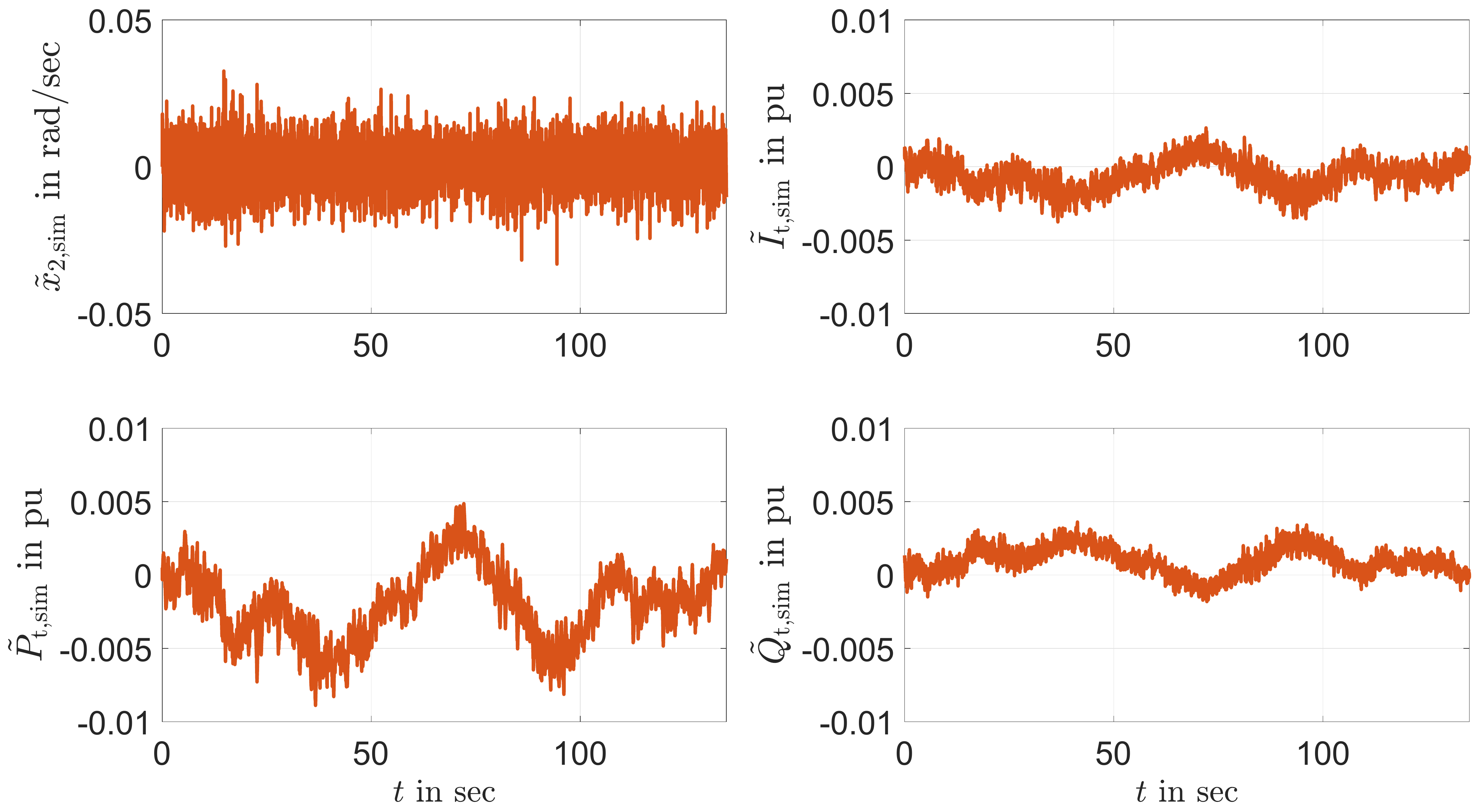}
	\caption{Error between the simulation results using the parameters obtained via the DREM-based parameter estimation and the mapped PMU-measurements for the cross-validation.}
	\label{fig:Crossvali_sim}
\end{figure}
\begin{table}
	\centering
		\begin{tabular}{l|l|l}
		\cmidrule{2-3}
		& State &sMAPE in \% \\ \cmidrule{1-3}
		Adaptive observer  & $x_{2}$ & 1.18~\% \\ \cmidrule{1-3}
		&   $x_{2}$ & 0.73~\% \\
		Simulation	 & $I_\mrm{t}$ & 0.05~\% \\	
		results	&  $P_\mrm{t}$& 0.22~\% \\	
		&  $Q_\mrm{t}$ & 0.07~\% \\	\cmidrule{1-3}
	\end{tabular}
	\caption{sMAPE of the adaptive observer and the simulation results for the cross-validation.}
	\label{tab:sMAPE_cross}
\end{table}
\section{Conclusions and future research}
\label{sec:conclusions}
The effectiveness of an algorithm for the problem of unknown-input DSE in multi-machine power systems was experimentally demonstrated using real-world PMU-measurements. The algorithm - originally presented by part of the authors in \cite{lorenz-meyer_pmu-based_2020-1} - was modified such that the measurement location, at a substation close to a power plant. Such measurement location is far more accessible to TSOs, than the actual terminal bus of the SG. For this, a mapping of the measurements from this substation to the terminal bus of the SG was derived. Moreover, the parameter estimator was extended to the case of time-varying and known mechanical torque $T_\mrm{m}$. This was possible, as a standard power control governor and turbine model used by 50Hertz Transmission GmbH shown in Figure \ref{fig:turbine_governor} is available and was used to generate $T_\mrm{m}$. Furthermore, the assumptions used in \cite{lorenz-meyer_pmu-based_2020-1} on the direct-axis transient reactance $x'_\mrm{d}$ and the quadrature-axis reactance $x_\mrm{q}$ being equal and the stator resistance $R_\mrm{s}$ being neglectable were dropped. Also, by utilizing a improved regressor extension, improved filtering with better attenuation of measurement noise and adaptive estimator gains, the algorithm showed significantly improved convergence and decreased excitation requirements when using the recorded real-world PMU-measurements. 
The algorithm was experimentally validated using real-world PMU-measurements acquired in cooperation with the German TSO 50Hertz Transmission GmbH and a PMU provided by Studio Elektronike Rijeka d.o.o. An auto- as well as a cross-validation was performed. The adaptive observer showed very good performance in both the auto- as well as the cross-validation. It showed low sMAPE values for the observed state $x_{2}$ and good estimates of the unknown parameters. Furthermore, in the event playback simulations for the auto- and cross-validation, the error between the mapped measurements and the simulated state $x_{2}^{\text{sim}}$, terminal active and reactive power $P_\mrm{t}^{\text{sim}}$, $Q_\mrm{t}^{\text{sim}}$ and current $I_\mrm{t}^{\text{sim}}$ showed low sMAPE values. 

In the future, the algorithm's accuracy might be improved by incorporating a more complex model of the auxiliary system of the power plant. Furthermore, the estimation of the parameters of the utilized, potentially simplified, governor and turbine model could be assumed unknown and included in the parameter estimation of the adaptive observer. Lastly, the SG model could be extended to the fourth order flux decay model. First results along this line are presented in \cite{bobtsov_state_2021}.

\appendix
\subsection{Proof of the algebraic observer}
\label{app:proofs}
We present the proof of Lemma \ref{pro1}.
\begin{proof} 
	Adding $(\mrm{j}x_\mrm{q}I_\mrm{d}-\mrm{j}x_\mrm{q}I_\mrm{d})\mrm{e}^{\mrm{j}(x_{1}-\frac{\pi}{2})}$ to \eqref{eq:stator_alg} and applying some algebraic manipulations gives
	\begin{equation*}
		\begin{split}
			\mrm{j} x_{3}\mrm{e}^{\mrm{j}(x_{1}-\frac{\pi}{2})} &= y_{1}\mrm{e}^{\mrm{j}y_{2}}+ R_\mrm{s} (I_\mrm{d} + \mrm{j} I_\mrm{q})\mrm{e}^{\mrm{j}(x_{1}-\frac{\pi}{2})}  + \\
			+( \mrm{j}x_\mrm{q}I_\mrm{d} &- \mrm{j}x_\mrm{q}I_\mrm{d}-x_\mrm{q}I_\mrm{q}   +x_\mrm{d}'I_\mrm{d})\mrm{e}^{\mrm{j}(x_{1}-\frac{\pi}{2})}  . 
		\end{split}	
	\end{equation*}
	By using $(I_\mrm{d} + \mrm{j} I_\mrm{q})\mrm{e}^{\mrm{j}(x_{1}-\frac{\pi}{2})} = y_{3} \mrm{e}^{\mrm{j}y_{4}} $ and further algebraic manipulations yields
	\begin{equation*}
		\begin{split}
			\left((x_\mrm{q} - x_\mrm{d}') I_\mrm{d} +  x_{3}\right) \mrm{e}^{\mrm{j}x_{1}} &= (R_\mrm{s} + \mrm{j} x_\mrm{q}) y_{3} \mrm{e}^{\mrm{j}y_{4}} +\\&+ y_{1} \mrm{e}^{\mrm{j}y_{2}} =  \psi.
		\end{split}	
	\end{equation*}
	As $\psi \in \mathbb{C}$, $x_{1}$ can be calculated as
	$$
	x_{1} = \arg\{\psi\}   .
	$$
	With $x_{1}$ known, $x_{3}$ is obtained as
	$$
	x_{3} = |\psi| - (x_\mrm{q}-x'_\mrm{d})\cos(\pi/2-x_{1}+y_{4})y_{3}  .$$
\end{proof}

\subsection{Practical modifications of the parameter estimator}
\label{app:para_est}
Following \cite{lorenz-meyer_pmu-based_2020-1}, we present details of the derivation of a DREM-based parameter estimator for $a_{1}$ and $a_{2}$ of the SG assuming $T_\mrm{m}$ time-varying and known, e.g., modeled via a governor and turbine model.

By using $x_{1}$ from \eqref{eq:alg_x1} and considering the swing equation \eqref{eq:x2}, we define the vector of unknown parameters
	\begequ
	\lab{the_append}
	\bm{\theta}\coloneqq \begin{bmatrix} a_{1}&a_{2}	\end{bmatrix}^\top,
	\endequ 
	the filter 
	\begin{equation*}
		\mathcal{F} = \frac{\lambda_1\lambda_2\lambda_3}{(\lambda_1+s)(\lambda_2+s)(\lambda_3+s)},
	\end{equation*}
	with tuning parameters $\lambda_j>0, \ j={1,2,3}$ and the signals 
	\begin{equation}
		\begin{split}
			\label{eq:regressor_non_ext}
			z & \coloneqq \mathcal{F}[s^2[x_{1}]], \\
			\bm{\psi}&\coloneqq \begmat{-\mathcal{F}[s[x_{1}]]\\ \mathcal{F}[T_\mrm{m}-T_\mrm{e}]},
		\end{split}
	\end{equation}
where $s$ denotes the Laplace operator.
	Thus, we can extend \eqref{eq:regressor_non_ext} with a linear, single-input 2-output, bounded-input bounded-output (BIBO)-stable operator $\bm{\mathcal{H}}$ and define the vector $\bm{Z} \in \rea^2$ and the matrix $\bm{\Psi} \in \rea^{2 \times 2}$
	\begin{equation}
		\begin{split}
			\label{eq:regressor_ext}
			\bm{Z} &\coloneqq \bm{\mathcal{H}}[z],\\
			\bm{\Psi} &\coloneqq \bm{\mathcal{H}}[(\bm{\psi})^\top],
		\end{split}
	\end{equation}
	the matrix $\bm{\mathcal{Z}} \in \rea^{2 \times 2}$ and the signal $\Delta$
	\begin{equation}
		\begin{split}
			\bm{\mathcal{Z}}&\coloneqq\adj\{\bm{\Psi}\} \bm{Z},\\
			\Delta & \coloneqq \det\{\bm{\Psi}\},
		\end{split}
	\end{equation}
	with $\det\{\cdot\}$ being the determinant and $\adj\{\cdot\}$ being the adjunct matrix. 
	Hence, taking the result in the time-domain the scalar parameter estimators can be defined as 
	\begin{equation}
		\label{eq:para_est}
	\dot {\hat \theta}_{j}=- \gamma_{j} K_j^\gamma  \Delta(\Delta  \hat \theta_{j} -\mathcal{Z}_{j}),\;j=1,2, 
	\end{equation}
	where $\gamma_j$ denotes the constant gain and $K_j^\gamma $ the additional time-varying gain defined as 
		\begin{equation}
		\label{eq:Kj}
		K^\gamma_{j} = \frac{\Delta^2_\mrm{ref}}{\bar{\Delta}^2},  
	\end{equation}
	where $\Delta^2_\mrm{ref}$ is the average value of $\Delta^2$ during a reference scenario, e.g. the auto-validation, and $\bar{\Delta}^2$ is the moving average of the $\Delta^2$.
	\\The dynamics of the parameter estimation error satisfies
	\begin{equation*}
	\dot {\tilde \theta}_{j}=\mrm{e}^{-\gamma_{j}\int_{0}^{t} K_j^\gamma \Delta^2 \mrm{d}\tau} \tilde{\theta}_{j}(0)  ,\;j=1,2,
	\end{equation*}
	where the parameter estimation error is defined as $\tilde \theta_{j}\coloneqq\hat \theta_{j} -\theta_{j}$.
	Provided that $K_j^\gamma$ is positive and lower and upper bounded, 
	the following inequality holds
	$$	100 \Delta^2\geq  K_j^\gamma\Delta^2\geq \epsilon\Delta^2, $$
	with $\epsilon>0$ and $100$ denoting the constant lower and upper bound of $K_j^\gamma$, respectively. Integration yields
	 \begin{equation*}
	 \begin{split}
	  100\liminf \int_0^t \Delta^2\mathrm{d}\tau  &\geq \liminf \int_0^t K_j^\gamma\Delta^2\mathrm{d}\tau \\ &\geq \epsilon\liminf \int_0^t \Delta^2\mathrm{d}\tau. \end{split}\end{equation*}
 	 Hence, it follows that 
  $$	\dot {\tilde \theta}_{j} \leq \mrm{e}^{-\gamma_{j}\epsilon\int_{0}^{t} \Delta^2 \mrm{d}\tau} \tilde{\theta}_{j}(0)  ,\;j=1,2,$$
  	and provided that $\gamma_{j}>0$ and $\Delta \notin \mathcal{L}_2$, that is
	  \begequ
	  \lab{eq:exccon}
	  \liminf \int_0^t \Delta^2\mathrm{d}\tau=\infty,
	  \endequ	
	 the parameter estimation error satisfies
	$$
	\liminf \tilde \theta_{j}(t)=0,\;j=1,2.
	$$

\bibliographystyle{IEEEtran}
\bibliography{Diss_bib}

\begin{thebibliography}{10}
\providecommand{\url}[1]{#1}
\csname url@samestyle\endcsname
\providecommand{\newblock}{\relax}
\providecommand{\bibinfo}[2]{#2}
\providecommand{\BIBentrySTDinterwordspacing}{\spaceskip=0pt\relax}
\providecommand{\BIBentryALTinterwordstretchfactor}{4}
\providecommand{\BIBentryALTinterwordspacing}{\spaceskip=\fontdimen2\font plus
\BIBentryALTinterwordstretchfactor\fontdimen3\font minus
  \fontdimen4\font\relax}
\providecommand{\BIBforeignlanguage}[2]{{%
\expandafter\ifx\csname l@#1\endcsname\relax
\typeout{** WARNING: IEEEtran.bst: No hyphenation pattern has been}%
\typeout{** loaded for the language `#1'. Using the pattern for}%
\typeout{** the default language instead.}%
\else
\language=\csname l@#1\endcsname
\fi
#2}}
\providecommand{\BIBdecl}{\relax}
\BIBdecl

\bibitem{winter_pushing_2015}
W.~Winter, K.~Elkington, G.~Bareux, and J.~Kostevc, ``Pushing the {Limits}:
  {Europe}'s {New} {Grid}: {Innovative} {Tools} to {Combat} {Transmission}
  {Bottlenecks} and {Reduced} {Inertia},'' \emph{IEEE Power and Energy
  Magazine}, vol.~13, no.~1, pp. 60--74, Jan. 2015, conference Name: IEEE Power
  and Energy Magazine.

\bibitem{milano_foundations_2018}
F.~Milano, F.~Doerfler, G.~Hug, D.~J. Hill, and G.~Verbic, ``Foundations and
  {Challenges} of {Low}-{Inertia} {Systems} ({Invited} {Paper}),'' in
  \emph{2018 {Power} {Systems} {Computation} {Conference} ({PSCC})}, Jun. 2018,
  pp. 1--25.

\bibitem{zhao_et_al_power_2021}
J.~Zhao~et al., ``Power {System} {Dynamic} {State} and {Parameter}
  {Estimation}-{Transition} to {Power} {Electronics}-{Dominated} {Clean}
  {Energy} {Systems},'' IEEE Power and Engineering Society, Tech. Rep., 2021.

\bibitem{vittal_et_al_next_2011}
V.~Vittal~et. al, ``Next {Generation} {On}-{Line} {Dynamic} {Security}
  {Assessment},'' PSERC Final Project Technical Report, Tech. Rep., 2011.

\bibitem{paul_dynamic_2018}
A.~Paul, G.~Joos, and I.~Kamwa, ``Dynamic {State} {Estimation} of {Full}
  {Power} {Plant} {Model} from {Terminal} {Phasor} {Measurements},'' in
  \emph{2018 {IEEE}/{PES} {Transmission} and {Distribution} {Conference} and
  {Exposition} ({T} {D})}, Apr. 2018, pp. 1--5, iSSN: 2160-8563.

\bibitem{wang_alternative_2012}
S.~Wang, W.~Gao, and A.~P.~S. Meliopoulos, ``An {Alternative} {Method} for
  {Power} {System} {Dynamic} {State} {Estimation} {Based} on {Unscented}
  {Transform},'' \emph{IEEE Transactions on Power Systems}, vol.~27, no.~2, pp.
  942--950, May 2012, conference Name: IEEE Transactions on Power Systems.

\bibitem{valverde_unscented_2010}
G.~Valverde and V.~Terzija, ``Unscented {Kalman} filter for power system
  dynamic state estimation,'' \emph{IET Generation, Transmission \&
  Distribution}, vol.~5, no.~1, pp. 29--37, 2010.

\bibitem{emami_particle_2015}
K.~Emami, T.~Fernando, H.~H.-C. Iu, H.~Trinh, and K.~P. Wong, ``Particle
  {Filter} {Approach} to {Dynamic} {State} {Estimation} of {Generators} in
  {Power} {Systems},'' \emph{IEEE Transactions on Power Systems}, vol.~30,
  no.~5, pp. 2665--2675, Sep. 2015, conference Name: IEEE Transactions on Power
  Systems.

\bibitem{cui_particle_2015}
Y.~Cui and R.~Kavasseri, ``A {Particle} {Filter} for {Dynamic} {State}
  {Estimation} in {Multi}-{Machine} {Systems} {With} {Detailed} {Models},''
  \emph{IEEE Transactions on Power Systems}, vol.~30, no.~6, pp. 3377--3385,
  Nov. 2015, conference Name: IEEE Transactions on Power Systems.

\bibitem{nugroho_robust_2020}
S.~A. Nugroho, A.~F. Taha, and J.~Qi, ``Robust {Dynamic} {State} {Estimation}
  of {Synchronous} {Machines} {With} {Asymptotic} {State} {Estimation} {Error}
  {Performance} {Guarantees},'' \emph{IEEE Transactions on Power Systems},
  vol.~35, no.~3, pp. 1923--1935, May 2020, conference Name: IEEE Transactions
  on Power Systems.

\bibitem{taha_risk_2018}
A.~F. Taha, J.~Qi, J.~Wang, and J.~H. Panchal, ``Risk {Mitigation} for
  {Dynamic} {State} {Estimation} {Against} {Cyber} {Attacks} and {Unknown}
  {Inputs},'' \emph{IEEE Transactions on Smart Grid}, vol.~9, no.~2, pp.
  886--899, Mar. 2018, conference Name: IEEE Transactions on Smart Grid.

\bibitem{qi_comparing_2018}
J.~Qi, A.~F. Taha, and J.~Wang, ``Comparing {Kalman} {Filters} and {Observers}
  for {Power} {System} {Dynamic} {State} {Estimation} {With} {Model}
  {Uncertainty} and {Malicious} {Cyber} {Attacks},'' \emph{IEEE Access},
  vol.~6, pp. 77\,155--77\,168, 2018, conference Name: IEEE Access.

\bibitem{simendic_-field_2005}
Z.~J. Simendic, C.~Vladimir, and G.~S. Svenda, ``In-field verification of the
  real-time distribution state estimation,'' in \emph{{CIRED} 2005 - 18th
  {International} {Conference} and {Exhibition} on {Electricity}
  {Distribution}}, 2005, pp. 1--4.

\bibitem{katic_field_2013}
N.~Katic, L.~Fei, G.~Svenda, and Z.~Yongji, ``Field testing of distribution
  state estimator,'' in \emph{22nd {International} {Conference} and
  {Exhibition} on {Electricity} {Distribution} ({CIRED} 2013)}, 2013, pp. 1--4.

\bibitem{lorenz-meyer_pmu-based_2020-1}
N.~Lorenz-Meyer, A.~Bobtsov, R.~Ortega, N.~Nikolaev, and J.~Schiffer,
  ``\BIBforeignlanguage{en}{{PMU}-based decentralised mixed algebraic and
  dynamic state observation in multi-machine power systems},''
  \emph{\BIBforeignlanguage{en}{IET Generation, Transmission \& Distribution}},
  vol.~14, no.~25, pp. 6267--6275, 2020.

\bibitem{ghahremani_local_2016}
E.~Ghahremani and I.~Kamwa, ``\BIBforeignlanguage{en}{Local and {Wide}-{Area}
  {PMU}-{Based} {Decentralized} {Dynamic} {State} {Estimation} in
  {Multi}-{Machine} {Power} {Systems}},'' \emph{\BIBforeignlanguage{en}{IEEE
  Transactions on Power Systems}}, vol.~31, no.~1, pp. 547--562, Jan. 2016.

\bibitem{van_cutsem_voltage_1998}
T.~Van~Cutsem and C.~Vournas, \emph{Voltage {Stability} of {Electric} {Power}
  {Systems}}.\hskip 1em plus 0.5em minus 0.4em\relax Boston: Springer, 1998.

\bibitem{sauer_power_2006}
P.~Sauer and M.~A. Pai, \emph{Power {Systems} {Dynamics} and
  {Stability}}.\hskip 1em plus 0.5em minus 0.4em\relax Wiley, 2006.

\bibitem{noauthor_ieee_2020}
``{IEEE} {Guide} for {Synchronous} {Generator} {Modeling} {Practices} and
  {Parameter} {Verification} with {Applications} in {Power} {System}
  {Stability} {Analyses},'' \emph{IEEE Std 1110-2019 (Revision of IEEE Std
  1110-2002)}, pp. 1--92, 2020.

\bibitem{machowski_power_2008}
J.~Machowski, B.~Janusz~W., and B.~James~R., \emph{Power {System} {Dynamics}:
  {Stability} and {Control}}, 2nd~ed.\hskip 1em plus 0.5em minus 0.4em\relax
  West Sussex, United Kingdom: John Wiley \& Sons, 2008.

\bibitem{glover_power_2012}
J.~D. Glover, S.~S. Mulukutla, and T.~J. Overbye, \emph{Power {System}
  {Analysis} and {Design}}, 5th~ed.\hskip 1em plus 0.5em minus 0.4em\relax
  Stamford, USA: Cengage Learning, 2012.

\bibitem{oeding_elektrische_2011}
D.~Oeding and B.~Oswald, \emph{Elektrische {Kraftwerke} und {Netze}},
  7th~ed.\hskip 1em plus 0.5em minus 0.4em\relax Berlin: Springer, 2011.

\bibitem{uecker_differential_2015}
L.~Uecker and K.~Wedeward, ``Differential flatness of the flux-decay generator
  model,'' in \emph{2015 10th {System} of {Systems} {Engineering} {Conference}
  ({SoSE})}, May 2015, pp. 146--151.

\bibitem{poznyak_advanced_2008}
A.~S. Poznyak, \emph{Advanced {Mathematical} {Tools} for {Automatic} {Control}
  {Engineers} - {Deterministic} {Techniques}}.\hskip 1em plus 0.5em minus
  0.4em\relax UK: Elsevier, 2008.

\bibitem{aranovskiy_performance_2017}
S.~Aranovskiy, A.~Bobtsov, R.~Ortega, and A.~Pyrkin, ``Performance
  {Enhancement} of {Parameter} {Estimators} via {Dynamic} {Regressor}
  {Extension} and {Mixing}*,'' \emph{IEEE Transactions on Automatic Control},
  vol.~62, no.~7, pp. 3546--3550, Jul. 2017, conference Name: IEEE Transactions
  on Automatic Control.

\bibitem{vidyasagar_decomposition_1980}
M.~Vidyasagar, ``Decomposition techniques for large-scale systems with
  nonadditive interactions: {Stability} and stabilizability,'' \emph{IEEE
  Transactions on Automatic Control}, vol.~25, no.~4, pp. 773--779, Aug. 1980,
  conference Name: IEEE Transactions on Automatic Control.

\bibitem{ramos_et_al_benchmark_2015}
R.~Ramos~et al., ``Benchmark {Systems} for {Small}-{Signal} {Stability}
  {Analysis} and {Control},'' IEEE PES Resource Center, Tech. Rep., 2015.

\bibitem{armstrong_long-range_1985}
J.~S. Armstrong, \emph{Long-{Range} {Forecasting}: from {Crystal} {Ball} to
  {Computer}}, 2nd~ed.\hskip 1em plus 0.5em minus 0.4em\relax New York: John
  Wiley \& Sons, 1985.

\bibitem{bobtsov_state_2021}
A.~Bobtsov, R.~Ortega, N.~Nikolaev, M.~N.~L. Lorenz-Meyer, and J.~Schiffer,
  ``State {Observation} of {Power} {Systems} {Equipped} with {Phasor}
  {Measurement} {Units}: {The} {Case} of {Fourth} {Order} {Flux}-{Decay}
  {Model},'' \emph{IEEE Transactions on Automatic Control}, pp. 1--1, 2021.

\end{thebibliography}

\begin{IEEEbiography}
	[{\includegraphics[width=1in,height=1.25in,clip,keepaspectratio]{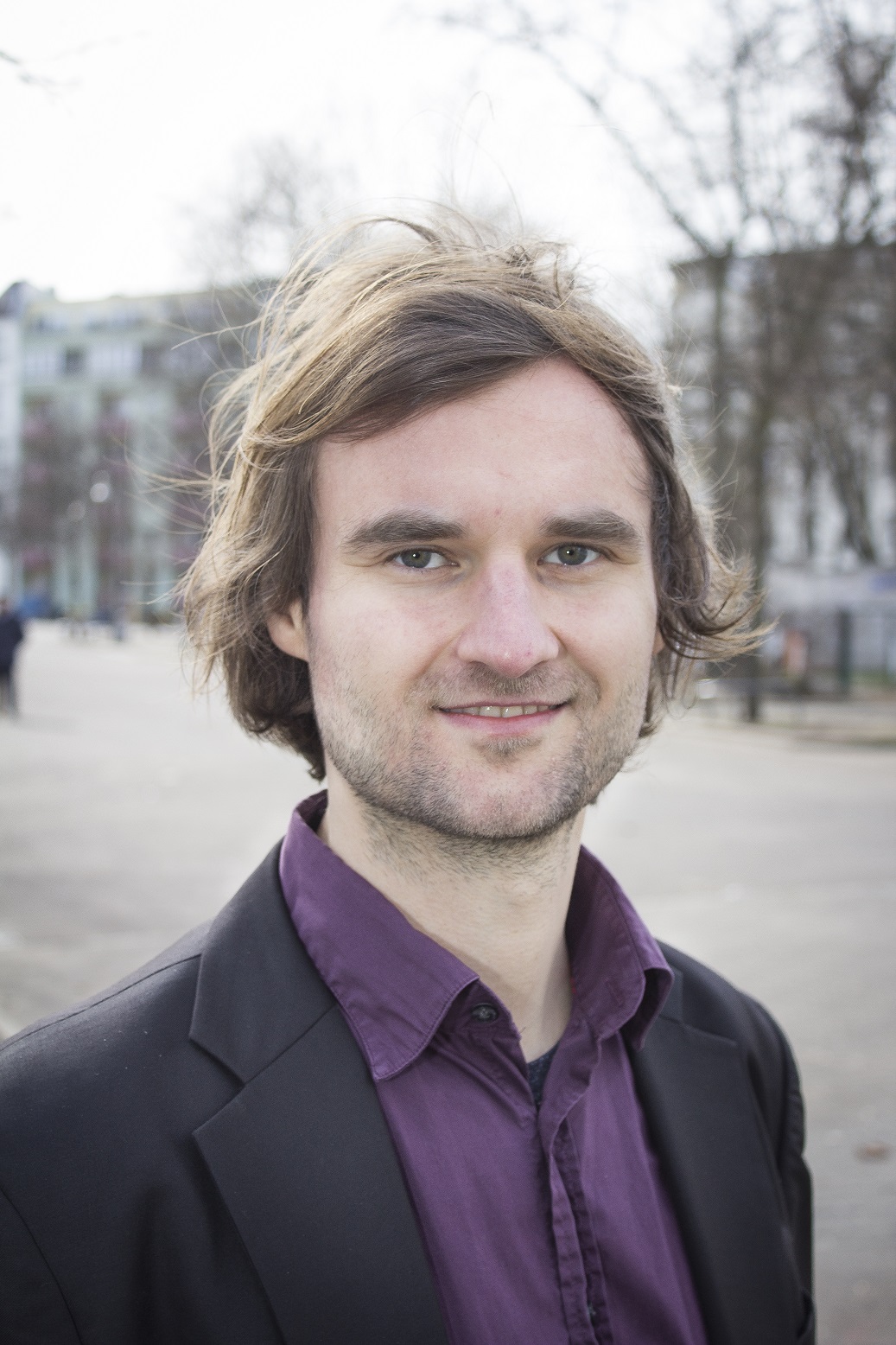}}]
	{Nicolai Lorenz-Meyer} received his M.Sc. in Engineering Science from the Technical University of Berlin in 2019 and his B.Eng. in Business Administration and Engineering for Environment and Sustainability from the Berlin University of Applied Sciences and Technology and the Berlin School of Economics and Law in 2016. 
	
	He is currently pursuing the Ph.D. degree with the the chair of Control Systems and Network Control Technology at the Brandenburg University of Technology Cottbus-Senftenberg, Germany. His current research interests include the development of control theory-based methods for on-line dynamics security assessment in power systems. 
\end{IEEEbiography}

\begin{IEEEbiography}
	[{\includegraphics[width=1in,height=1.25in,clip,keepaspectratio]{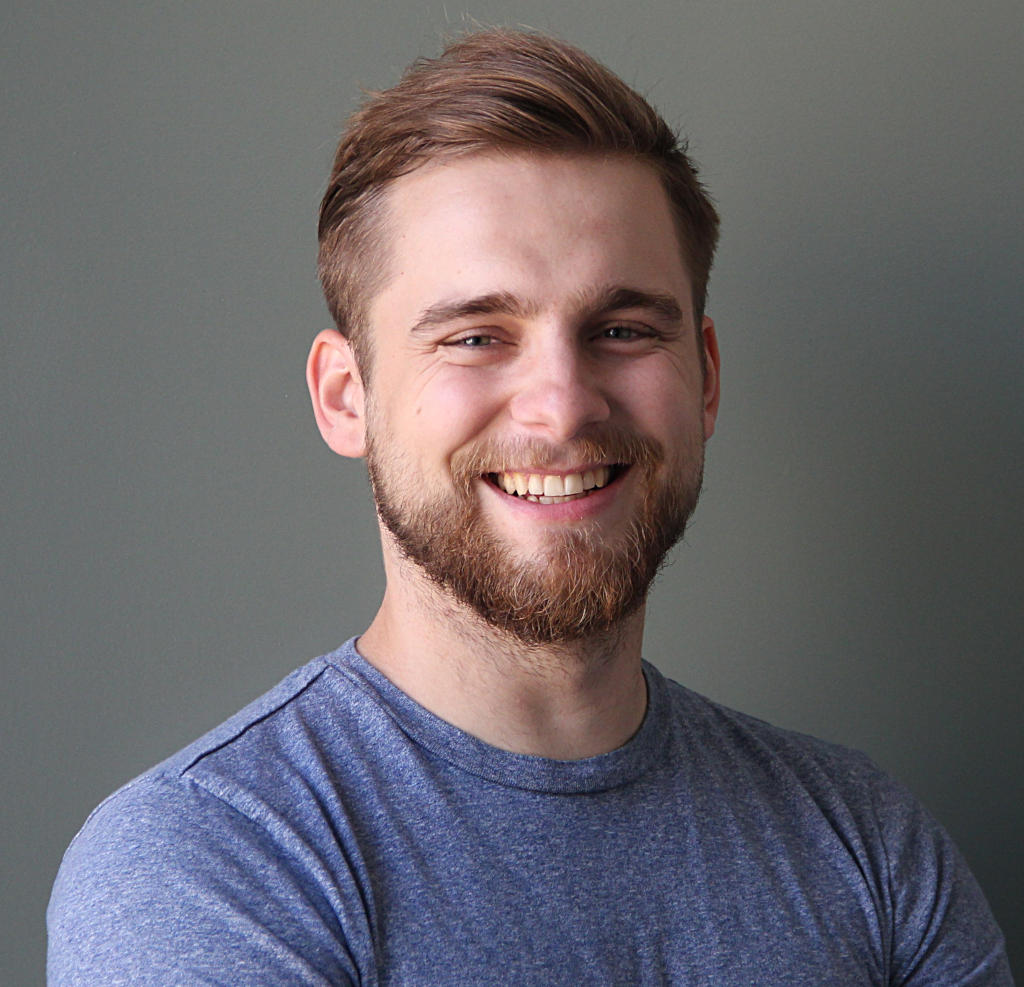}}]
	{Ren\'e Suchantke} was born in Berlin, Germany in 1990. He received the B.Eng. at \textit{Beuth University of Applied Sciences} in 2013, his M.Sc. and Ph.D. degrees at the High Voltage Department at \textit{TU Berlin} in 2014 and 2018, respectively. Other research projects included power network simulations in \textit{EMTP-RV} and different FEM simulations with \textit{Comsol Multiphysics}. Since 2019 he is with the grid planning department of German TSO \textit{50Hertz Transmission GmbH}.
\end{IEEEbiography}

\begin{IEEEbiography}
	[{\includegraphics[width=1in,height=1.25in,clip,keepaspectratio]{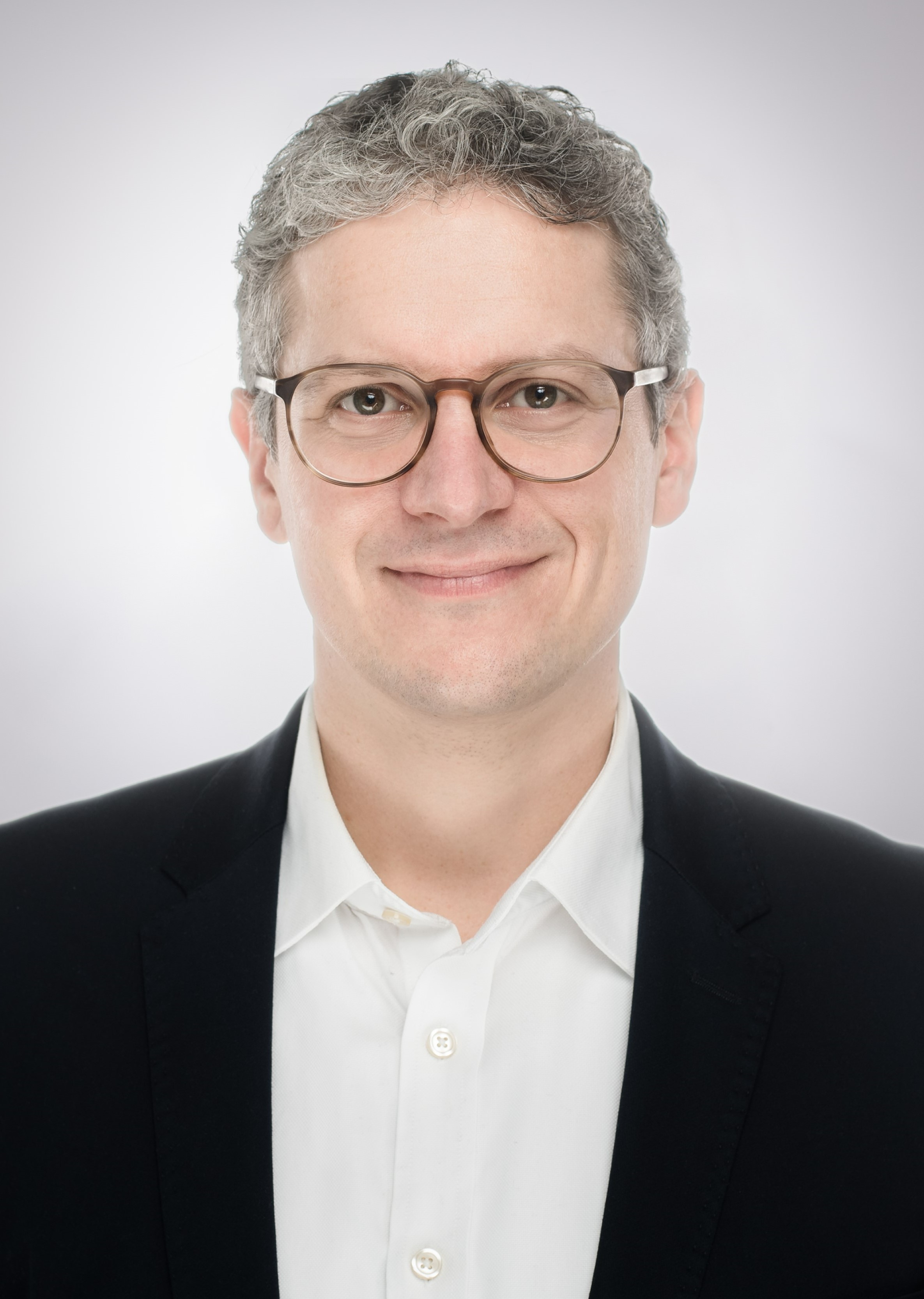}}]
	{Johannes Schiffer} received the Diploma degree in engineering cybernetics from the University of Stuttgart, Germany, in 2009 and the Ph.D. degree (Dr.-Ing.) in electrical engineering from Technische Universit\"at (TU) Berlin, Germany, in 2015. 
	
	He currently holds the chair of Control Systems and Network Control Technology at Brandenburgische Technische Universit\"at Cottbus-Senftenberg, Germany and leads the business area Control, Automation and Operation Managemement at the Fraunhofer Research Institution for Energy Infrastructures and Geothermal Systems (IEG). Prior to that, he has held appointments as Lecturer (Assistant Professor) at the School of Electronic and Electrical Engineering, University of Leeds, U.K. and as Research Associate in the Control Systems Group and at the Chair of Sustainable Electric Networks and Sources of Energy both at TU Berlin.
	
	In 2017 he and his co-workers received the Automatica Paper Prize over the years 2014-2016. His current research interests include distributed control and analysis of complex networks with application to microgrids and smart energy systems.
\end{IEEEbiography}

\end{document}